\documentclass{jfm}

\graphicspath{{figures/}}
\usepackage{graphicx}
\usepackage{natbib}
\usepackage{hyperref}
\usepackage{amsmath}
\usepackage{color}
\usepackage{pgfplots}

\newcommand{\vect}[1]{\boldsymbol{#1}}
\newcommand{\aver}[1]{ \! \left\langle {#1} \right \rangle \!}

\title{Curvature effects on the structure of near-wall turbulence}
\author[D. Selvatici, M. Quadrio, A. Chiarini]
{Davide Selvatici \corresp{Current address: Physics of Fluids Group, Max Planck Center Twente for Complex Fluid Dynamics, J. M. Burgers Center for Fluid Dynamics, University of Twente, P. O. Box 217, Enschede 7500 AE, Netherlands},
Maurizio Quadrio
and Alessandro Chiarini \corresp{Current address: Complex Fluids and Flows Unit, Okinawa Institute of Science and Technology Graduate University, 1919-1 Tancha, Onna-son, Okinawa 904-0495, Japan} \corresp{\email{alessandro.chiarini@oist.jp}}}

\affiliation{Dipartimento di Scienze e Tecnologie Aerospaziali, Politecnico di Milano, via La Masa 34, 20156 Milano, Italy
}

\begin{document}

\maketitle
	
\begin{abstract}
The interaction between near-wall turbulence and wall curvature is described for the incompressible flow in a plane channel with a small concave-convex-concave bump on the bottom wall, with height comparable to the wall-normal location of the main turbulent structures. The analysis starts from a database generated by a direct numerical simulation and hinges upon the anisotropic generalised Kolmogorov equations, i.e. the exact budget equations for the second-order structure function tensor. The influence of the bump on the wall cycle and on the energy production, redistribution and transfers is described in the physical and scale spaces.
%
Over the upstream side of the bump, the energy drained from the mean flow to sustain the streamwise fluctuations decreases, and the streaks of high and low streamwise velocity weaken and are stretched spanwise. 
After the bump tip, instead, the production of streamwise fluctuations grows and the streaks intensify, progressively recovering their characteristic spanwise scale.
The wall-normal fluctuations, and thus the quasi-streamwise vortices, are sustained by the mean flow over the upstream side of the bump, while energy flows from the vertical fluctuations to the mean field over the downstream side.
On the concave portion of the upstream side, the near-wall fluctuations form structures of spanwise velocity which are consistent with Taylor--G\"ortler vortices at an early stage of development. Their evolution is described by analysing the scale-space pressure--strain term.
A schematic description of the bump flow is presented, in which various regions are identified according to the signs of curvature and streamwise pressure gradient.
\end{abstract}

\section{Introduction} 
\label{sec:introduction}

Unveiling the dynamics of turbulent fluctuations, their sustaining mechanism and their tendency to organise into coherent structures, i.e. regions that exhibit significant correlation over a range of space and/or time larger than the smallest scale of the flow \citep{robinson-1991b}, has interested scholars for decades: a full understanding would improve our ability to predict the mean flow and, eventually, to control it.
A large number of studies has been devoted to identify and extract coherent structures in flows bounded by planar walls, such as channel flows and boundary layers, and to characterise their role in the sustaining mechanism of turbulence \citep{blackwelder-kaplan-1976, robinson-1991b, hamilton-kim-waleffe-1995, panton-1997, jimenez-pinelli-1999, schoppa-hussain-2002}. Streaks of high and low streamwise velocity (HSS and LSS) and quasi-streamwise vortical structures (QSVs) are known to be major players in the self-sustained near-wall cycle; see \cite{jeong-etal-1997} for an overview and \cite{jimenez-2022} for a more recent discussion. These coherent structures are dynamically related, as the QSVs advect higher/lower momentum downwards/upwards, generating HSS and LSS.

A plane wall, however, is just a simplified setting. In real-world applications, a turbulent flow often develops over curved walls, which may induce favourable or adverse pressure gradients, a non-constant friction along the flow direction and, possibly, flow separation. 
In this work, we consider walls with mild, localised curvature, and aim at enriching the canonical description of the near-wall turbulent cycle with an extensive scale-space analysis of curvature effects. The simple flow chosen for the study develops in a plane channel, where one of the two walls has a small bump.

The turbulent flow over a bump has been extensively studied, both experimentally and numerically. It is used as a test case for developement and validation of LES and RANS models \citep{wu-squires-1998, frohlich-etal-2005, breuer-etal-2009} and for exploring the dynamics of recirculating regions in both the laminar \citep{gallaire-etal-2007} and turbulent \citep{mollicone-etal-2018} regimes. 
\cite{baskaran-smits-joubert-1987} and \cite{webster-degraaff-eaton-1996-1, webster-degraaff-eaton-1996-2} experimentally investigated the development of a turbulent boundary layer over a concave-convex-concave bump which does not produce flow separation, and found that the mean velocity profile over the bump significantly deviates from the law of the wall. 
Moreover, as confirmed later numerically \citep{wu-squires-1998}, the concave-convex change of curvature on the upstream side of the bump and the convex-concave one on the downstream side trigger two internal layers; the upstream one rapidly grows far from the wall owing to the locally adverse pressure gradient. 
\cite{marquillie-laval-dolganov-2008} studied via direct numerical simulation (DNS) the budget equation for the turbulent kinetic energy in a channel flow with a bump that produces a small recirculation, and assessed the effect of the curvature on the velocity streaks by resorting to velocity correlations. 
They found that strong coherent structures form near the separated region, and that the LSS and HSS are first stretched in the spanwise direction by the favourable pressure gradient and subsequently disappear after the flow separation on the downstream side of the bump. 
Later, \cite{marquillie-ehrenstein-laval-2011} considered a larger value of the Reynolds number and focused on the dynamics of the LSS, observing that the onset of their instability coincides with a strong production of turbulent kinetic energy. 
\cite{mollicone-etal-2017} studied via DNS the dynamics of the separation bubble over the downstream side of a convex bump of parabolic shape, considering various bump heights and Reynolds numbers. 
A larger height was found to lead to a larger separation bubble, while a larger Reynolds number implies a smaller bubble and a shear layer that remains closer to the wall. 
They used the budget equation for the turbulent kinetic energy to describe the production, transfer and dissipation of fluctuating energy in physical space. Turbulent kinetic energy, after being produced in the separating shear layer due to the pressure drop, is transferred within the bubble and downstream of the bump, where it is eventually dissipated.
In a later work, the same authors \citep{mollicone-etal-2018} provided a dynamical description of turbulence in the recirculating region behind the same bump using the generalised Kolmogorov equation (GKE). 
The GKE \citep{hill-2001, danaila-etal-2001} is the exact budget equation for the second-order structure function, which is commonly interpreted as scale energy. As such, the GKE describes production, transport and dissipation of turbulent energy, considering simultaneously the physical space and the space of scales. 
\cite{mollicone-etal-2018} found that the turbulent energy is mainly produced at specific scales in the shear layer, and that its transfer towards the sink flow regions occurs by means of both inverse and direct energy cascades.

Most of the work mentioned above considered relatively large bumps that drastically alter the structure of the flow and produce a massive separation, and studied the dynamics of the recirculating bubble. Smaller bumps have received less attention, and the description of how a spatially localised curvature of the wall, mild enough to avoid recirculation, affects the structure of the flow is lacking. This is the goal of the present work. 

\cite{moser-moin-1987} were first to consider with DNS how curvature affects a fully developed turbulent duct flow. In that case, however, the streamwise direction is homogeneous, and the curvature is constant with the streamwise coordinate. 
Over concave walls, new structures arise and alter the profiles of the Reynolds stresses. These structures are referred to as Taylor--G\"ortler vortices (TGVs), first detected by \cite{gortler-1941} in laminar boundary layers over curved walls. They are pairs of counter-rotating and streamwise-aligned vortices, which originate because of an imbalance between the centrifugal force and the radial pressure gradient. 
For details on the TGVs and the related instability, we refer the interested reader to \cite{floryan-1991, saric-1994, luchini-bottaro-1998} and to the more recent works of \cite{xu-liu-wu-2020} and \cite{dagaut-etal-2021}. 
Whether TGVs are present in turbulent flows with localised curvature is still unclear. In particular, TGVs have never been observed in the turbulent bump flow. For example, \cite{webster-degraaff-eaton-1996-1} explicitly state that such vortices are absent in the flow over the bump they considered, because the concave region is too short. 
\cite{baskaran-smits-joubert-1987} experimentally studied the response of a turbulent boundary layer to sudden changes of the wall curvature and pressure gradient, considering a concave-convex-concave bump. In the region of concave curvature they did not find evidence of TGVs, and asked the question whether a convex curvature following a concave one can suppress these vortices. 
\cite{hall-1985} considered a three-dimensional laminar boundary layer and found that, in this case, TGVs are suppressed because of a cross-flow instability mode prevailing over the TGVs instability mode. 
\cite{benmalek-saric-1994} studied the non linear evolution of TGVs over a wall of variable curvature by means of the parabolised disturbance equations. They found that a convex curvature significantly stabilises disturbances introduced by an upstream concave region, resulting into a decay of the TGVs. 
\cite{xu-liu-wu-2020} investigated the TGVs in a laminar boundary layer over a concave wall in a contracting/expanding stream that generates an adverse/favourable pressure gradient, at a G\"ortler number of order 1. They found that in presence of an adverse/favourable pressure gradient the TGVs saturate earlier/later and at a lower/higher amplitude, if compared to the zero-pressure-gradient case. 
To the best of our knowledge, evidence of TGVs in a fully developed turbulent flow over a surface with a localised concave-convex change of curvature is not available.

In this work we investigate the turbulent flow past a small concave-convex-concave bump that, unlike bumps considered in previous studies, does not generate a strong recirculation, and does not disrupt the upstream near-wall flow. Our specific aim is to provide an extensive description of the effect of mild and localised changes of the wall curvature on the organisation of the near-wall turbulence, and on the sustaining mechanism of the velocity fluctuations. We start from the DNS database produced by \cite{banchetti-luchini-quadrio-2020}, and study it by leveraging the anisotropic generalised Kolmogorov equations, or AGKE \citep{gatti-etal-2020}, to deal simultaneously with scales and positions. The AGKE are a set of exact budget equations for each component of the second-order structure function tensor, and describe production, transport, redistribution and dissipation of the Reynolds stresses in the combined space of scales and positions. Unlike the GKE used by \cite{mollicone-etal-2018}, the AGKE include a pressure--strain term that is essential to describe energy redistribution among components. Interesting insights in the flow physics will be obtained, by describing how curvature affects the scale-space energy production, redistribution and transfers. Moreover, statistical evidence of TGVs will be provided, and their evolution will be described.


The paper is organised as follows. In \S\ref{sec:DNS} the DNS database is described, while in \S\ref{sec:AGKE} the AGKE are briefly recalled. Section \S\ref{sec:flow-topology} deals with the topology of the mean flow, and identifies the main regions of the bump flow. Then in \S\ref{sec:before} and \S\ref{sec:after} the AGKE analysis is presented separately for the upstream and downstream parts of the bump. In \S\ref{sec:fluxes} the effect of the bump on the scale-space energy transfers is studied. Lastly, \S\ref{sec:conclusions} provides some concluding remarks. The statistical results obtained in the present work are made available in the public repository at \url{https://doi.org/10.5281/zenodo.7879911}.

\section{Methods}
\label{sec:methods}
 
\subsection{The DNS database} 
\label{sec:DNS}
\begin{figure}
\centering
\includegraphics[width=\linewidth]{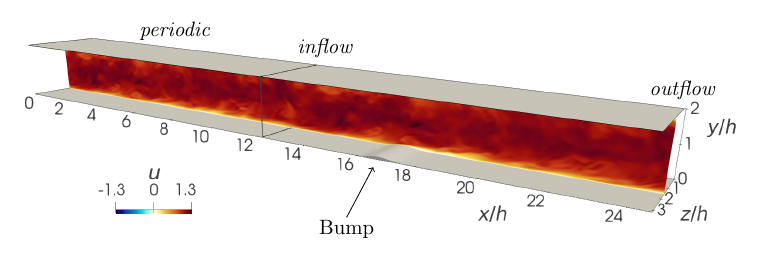}
\caption{Sketch of the computational domain and the reference frame. The portion with $x/h < 4 \pi$ corresponds to a periodic plane channel flow; the peak of the small bump is at $x/h \approx 16.7$. An instantaneous snapshot of the streamwise velocity field $u$ is plotted on a streamwise-parallel plane; flow is from left to right.}
\label{fig:geometry}
\end{figure}
We use the DNS database of the turbulent flow over a small bump produced by \cite{banchetti-luchini-quadrio-2020}. Their paper contains full details on the numerical method and the related computational procedures, which are only briefly recalled here. 
Figure \ref{fig:geometry} sketches the computational domain and the reference frame. A Cartesian coordinate system is used with the $x$, $y$ and $z$ axes respectively denoting the streamwise, wall-normal and spanwise directions.
The computational domain consists of two streamwise-adjacent blocks. The upstream portion is a canonical periodic channel with planar walls, and its streamwise length is $4 \pi h$, where $h$ is half the channel height. The downstream portion, with the bump on the lower wall, extends from $x=4 \pi h$ (where an inflow boundary condition is used) to $x= 4 \pi h + 12 h$ (where an outflow boundary condition is used), with dimensions $(L_x,L_y,L_z)=(12h,2h,2\pi h)$. The bump is two-dimensional, and its geometry is analytically defined by the sum of two overlapping Gaussian curves, i.e.
\begin{equation}
\frac{H(x)}{h} = a \exp \left[ - \left( \frac{x/h-4\pi-b}{c} \right)^2 \right] + a' \exp \left[ - \left( \frac{x/h-4\pi-b'}{c'} \right)^2 \right],
\label{eq:bump}
\end{equation}
with $a=0.0505$, $b=4$, $c=0.2922$ and $a'=0.060425$, $b'=4.36$, $c'=0.3847$. The maximum height of the bump is $h_b=0.0837h$ at $x \approx 16.7h$. This geometry resembles that in \cite{marquillie-laval-dolganov-2008}, but has a significantly smaller size to reduce blockage and to decrease the size of the recirculating region over the downhill side. 

The flow is governed by the incompressible Navier--Stokes equations for velocity $\vect{u}=(u,v,w)$ and pressure $p$. No-slip and no-penetration conditions are applied at the walls, and periodic conditions are set at the spanwise boundaries. In the upstream portion, periodic conditions are used also in the streamwise direction, whereas inflow and convective outflow conditions are used for the downstream portion.

The bulk Reynolds number $Re_b=U_b h /\nu$, based on the bulk velocity $U_b$ and on the channel half-height $h$, is $Re_b=3173$; $\nu$ is the kinematic viscosity of the fluid. For the upstream plane channel flow, it corresponds to a friction Reynolds number of $Re_\tau=u_\tau h / \nu=200$, where $u_\tau=\sqrt{\tau_w/\rho}$ is the friction velocity defined with the wall shear stress $\tau_w$ and density $\rho$.

The Navier--Stokes equations are integrated in time using the DNS code introduced by \cite{luchini-2016}, written in the CPL computer programming language \citep{cpl-website, luchini-2021}. The code uses second-order accurate finite differences on a staggered grid in the three directions. An implicit, second-order accurate immersed-boundary method is used to deal with the non-planar wall \citep{luchini-2013}. The computational domain is discretised with $(N_x,N_y,N_z)=(1120,312,241)$ points. In the wall-normal and spanwise directions, the same distribution of points is adopted in the upstream and downstream portions of the domain, to avoid interpolation. In the spanwise direction the distribution is uniform, yielding a resolution of $\Delta z \approx 0.01h$ that in the upstream portion corresponds to $\Delta z^+ \approx 2$ (viscous or plus units are defined with the local friction velocity $u_\tau$). In the wall-normal direction, instead, the resolution is higher close to the wall, with the grid spacing in the vicinity of the bump being $\Delta y^+ \approx 0.2$. For the streamwise direction, an uniform distribution is used in the upstream channel leading to $\Delta x \approx 0.01h $ and $\Delta x^+ \approx 2$. In the downstream portion, instead, a larger number of points is employed close to the bump to maintain the same resolution in viscous units, $\Delta x^+ \approx 2 $, despite the variation of the local friction.

The momentum equation is advanced in time by a fractional-step method using a third-order Runge--Kutta scheme. 
The Poisson equation for the pressure is solved using an iterative SOR algorithm. The time step is set at $\Delta t = 1.5 \times 10^{-3} h/U_b$, corresponding to an averaged Courant-Frederic-Levy (CFL) number of approximately $0.5$. 
After reaching statistical equilibrium, statistics are accumulated for $T \approx 1000 \ h/U_b$, with the database consisting of $335$ equally spaced snapshots.

Unless otherwise indicated, hereafter quantities are made dimensionless with $h$ and $U_b$; capital letters refer to mean fields, while small letters indicate the fluctuations around them. 
Throughout the paper, $\Delta y$ is used to indicate the distance from the lower wall in the $y$ direction; given the small slope of the bump, $\Delta y$ remains a good proxy for the actual wall distance even on the curved portion of the surface, with a difference that is everywhere less than 0.4\%.

\subsection{The anisotropic generalised Kolmogorov equations}
\label{sec:AGKE}

The anisotropic generalised Kolmogorov equations (AGKE) are exact budget equations for the second-order structure function tensor $\aver{\delta u_i \delta u_j}$, derived without approximations by manipulation of the Navier--Stokes equations \citep[see][for full details]{gatti-etal-2020}. The operator $\aver{\cdot}$ indicates ensemble averaging as well as averaging in time, if the flow is statistically stationary, and in the homogeneous directions. The AGKE are useful to study anisotropic, inhomogeneous and multiscale turbulent flows as they provide a dynamical description of turbulence considering simultaneously the physical space and the space of scales. The structure function tensor $\aver{\delta u_i \delta u_j}$ is based on the increment of the fluctuating velocity vector $\delta \vect{u} = \vect{u}(\vect{x}_b)-\vect{u}(\vect{x}_a)$ between two points $\vect{x}_b$ and $\vect{x}_a$, with $\vect{X}=(\vect{x}_a + \vect{x}_b)/2$ and $\vect{r}=\vect{x}_b-\vect{x}_a$ being their mid point and separation vector. In the general case, $\aver{\delta u_i \delta u_j}$ is a function of seven independent variables, i.e. the six coordinates of $\vect{X}$ and $\vect{r}$, and time $t$. 
In the present case, the AGKE are applied to a statistically stationary flow, with the spanwise $z$ direction being statistically homogeneous. Therefore the independent variables reduce to five, i.e. $(r_x,r_y,r_z,X,Y)$. The tensor $\aver{\delta u_i \delta u_j}$ incorporates the covariances $\aver{u_i u_j}$ of the velocity fluctuations and the spatial correlation tensor $R_{ij}$:
\begin{equation}
  \aver{\delta u_i \delta u_j} (\vect{X},\vect{r}) = V_{ij}(\vect{X},\vect{r}) - R_{ij}(\vect{X},\vect{r}) - R_{ji}(\vect{X},\vect{r})
\end{equation} 
where
\begin{equation}
  V_{ij}(\vect{X},\vect{r}) = \aver{u_i u_j}\left(\vect{X}+\frac{\vect{r}}{2}\right) + \aver{u_i u_j}\left(\vect{X}-\frac{\vect{r}}{2}\right)
  \label{eq:dudu-corr}
\end{equation}
is the sum of the covariances evaluated at $\vect{X} \pm \vect{r}/2$, and
\begin{equation}
  R_{ij}(\vect{X},\vect{r}) = \aver{u_i \left( \vect{X} + \frac{\vect{r}}{2} \right) u_j \left( \vect{X} - \frac{\vect{r}}{2} \right) }.
\end{equation}
For large enough $|\vect{r}|$ the correlation tensor vanishes, and $\aver{\delta u_i \delta u_j}$ reduces to $V_{ij}$. At large separations, thus, the AGKE are equivalent to the sum of the budget equations for the single-point Reynolds stresses evaluated at $\vect{X} \pm \vect{r}/2$. Therefore, the combination of the AGKE and the budget for the Reynolds stresses is equivalent to the K\'arm\'an--Howart equation for the correlation tensor. Thus, the information provided by the AGKE is equivalent to the more classical description based on the spectrally-decomposed Reynolds stress budgets \citep{kawata-alfredsson-2018,lee-moser-2019}, whenever the latter is feasible. Moreover, in the limiting case of stationary, homogeneous, isotropic turbulence the AGKE reduces to the classical Kolmogorov equation where the only independent variable is $r = |\vect{r}|$.

The AGKE, once tailored to the present flow, can be compactly written as:
\begin{equation}
  \frac{\partial \phi_{k,ij}}{\partial r_k} + \frac{\partial \psi_{\ell,ij}}{\partial X_\ell} = \xi_{ij}.
\label{eq:AGKE}
\end{equation}
The l.h.s. of equation \eqref{eq:AGKE} features the divergence of the flux vector $\vect{\Phi}_{ij}=(\vect{\phi}_{ij},\vect{\psi}_{ij})$, where the scale flux $\vect{\phi}_{ij}$ and the spatial flux $\vect{\psi}_{ij}$ read:
\begin{equation}
\phi_{k, ij} = \underbrace{\left \langle \delta U_k \delta u_i \delta u_j \right \rangle}_{\textrm{mean transport}}\ + \underbrace{\left \langle \delta u_k \delta u_i \delta u_j \right \rangle}_{\textrm{turbulent transport}}\ \underbrace{-\ 2 \nu \frac{\partial}{\partial r_k}\left \langle \delta u_i \delta u_j \right \rangle}_{\textrm{viscous diffusion}} \ \ k=1,2,3;
\label{eq:phi}
\end{equation}
\begin{equation}
\begin{gathered}
\psi_{\ell, ij} = 
 \underbrace{  \aver{ U_\ell^* \delta u_i \delta u_j }}_{\textrm{mean transport}} 
 + \underbrace{ \aver{ u_\ell^* \delta u_i \delta u_j }}_{\textrm{turbulent transport}}
 +\underbrace{ \frac{1}{\rho} \aver{ \delta p \delta u_i } \delta_{\ell j} 
              +\frac{1}{\rho} \aver{ \delta p \delta u_j } \delta_{\ell i}}_{\textrm{pressure transport}} + \\
\ \underbrace{- \frac{\nu}{2}\frac{\partial}{\partial X_\ell} \aver{ \delta u_i \delta u_j }}_{\textrm{viscous diffusion}} \ \ \ell=1,2;
\end{gathered}
\label{eq:psi}
\end{equation}
here $\delta_{ij}$ is the Kronecker delta. 
In each flux the mean, turbulent and pressure transports and the viscous diffusion are recognised, in analogy with the usual decomposition of the single-point budget equations for the Reynolds stresses \citep{pope-2000}.
The r.h.s. of equation \eqref{eq:AGKE}, $\xi_{ij}$, is the source term and is defined as:
\begin{equation}
\begin{gathered}
\xi_{ij} =
 \underbrace{ -\aver{ u_k^* \delta u_j} \delta \left( \frac{\partial U_i}{\partial x_k} \right) 
              -\aver{ u_k^* \delta u_i} \delta \left( \frac{\partial U_j}{\partial x_k} \right) 
              -\aver{ \delta u_k \delta u_j } \left( \frac{\partial U_i}{\partial x_k} \right)^*
              -\aver{ \delta u_k \delta u_i } \left( \frac{\partial U_j}{\partial x_k} \right)^* }_{\displaystyle P_{ij}} +\\
 \underbrace{ + \frac{1}{\rho} \aver{ \delta p \frac{\partial \delta u_i}{\partial X_j} } 
              + \frac{1}{\rho} \aver{ \delta p \frac{\partial \delta u_j}{\partial X_i} }        }_{\displaystyle \Pi_{ij}}
 \ - \underbrace{ 4 \epsilon_{ij}^*}_{\displaystyle D_{ij}};
\end{gathered}
\label{eq:source}
\end{equation}
the superscript $*$ indicates the average between the two points $\vect{X} \pm \vect{r}/2$. The source $\xi_{ij}(\vect{X},\vect{r})$ describes the net production of $\aver{\delta u_i \delta u_j}$ in the space of scales ($\vect{r}$) and in the physical space ($\vect{X}$). $P_{ij}$ is the production term and describes the energy exchange between the mean field and the fluctuating field. $\Pi_{ij}$ is the pressure--strain term and describes the energy redistribution among the components of $\aver{\delta u_i \delta u_j}$. $D_{ij}$ is the viscous dissipation, and is defined via the pseudo-dissipation tensor $\epsilon_{ij}=\nu \aver{ (\partial u_i / \partial x_k) (\partial u_j / \partial x_k)}$. The sum of the AGKE for the three diagonal components yields the generalised Kolmogorov equation (GKE) for the turbulent kinetic energy \citep{hill-2001,danaila-etal-2001}. 


Similarly to the budget equation for the Reynolds stress tensor (here recovered for large separations), designating fluxes and sources is partly conventional, with the implication that cause-effect relationships between them must be considered with care. The pressure transport term, indeed, may be moved at the right hand side and combined with the pressure--strain term to obtain the velocity--pressure gradient term $R_{ij} = \aver{\delta u_i \partial \delta p / \partial X_j} + \aver{\delta u_j \partial \delta p / \partial X_i}$. We prefer to keep a separate pressure--strain term as it reveals the redistributive effect of the fluctuating pressure: owing to continuity, $\Pi_{ij}$ vanishes for the budget of the scale energy (trace of the $\aver{\delta u_i \delta u_j}$ tensor). 
Also, one may or may not combine viscous diffusion with pseudo-dissipation, to obtain dissipation \citep{pope-2000}. 
It is worth mentioning that, as stated in \cite{jimenez-2016} for the budget of the Reynolds stresses, the AGKE is singular when interpreted as an equation for the vector of the fluxes $(\vect{\psi}_{ij},\vect{\phi}_{ij})$. Indeed, its solution is defined up to a solenoidal vector field. Therefore, one should consider several fluxes that differ for a zero-divergence term, and discriminate properties intrinsic to the flow physics from those linked to the particular choice of the flux vector. However, this aspect is beyond the scope of the present work, and we focus only on the vector of the fluxes as obtained directly after manipulation of the Navier--Stokes equations.

The AGKE have been successfully used \citep{chiarini-etal-2021} in a turbulent Couette flow to describe the role of the main structures in the sustaining mechanism of the velocity fluctuations, and to reveal the coexistence of direct and inverse cascades of turbulent stresses; they have been also employed to describe the structure of turbulence in the flow around an elongated rectangular cylinder \citep{chiarini-etal-2022}. The AGKE terms are computed here with the same code developed by \cite{gatti-etal-2019} and used by \cite{chiarini-etal-2022}, which -- for computational efficiency -- evaluates correlations pseudospectrally whenever possible, leveraging the Parseval's theorem. 

A preliminary study of the Reynolds stress tensor budget equations in the whole domain allows us to restrict the discussion of the AGKE terms to the subdomain $15 \le x \le 22$, $ 0 \le y \le 0.8$ centered on the bump.
We will consider the $r_x=r_y=0$ subspace. Indeed, the near-wall structures are known to be mostly aligned in the streamwise direction, hence their statistical trace is best seen considering spanwise and wall-normal separations \citep{gatti-etal-2020}. The analysis of the $r_x \neq 0$ or $r_y \neq 0$ subspaces (not shown) does not provide additional insight.

\section{Flow topology}
\label{sec:flow-topology}
\begin{figure}
\centering
\includegraphics[width=1\textwidth]{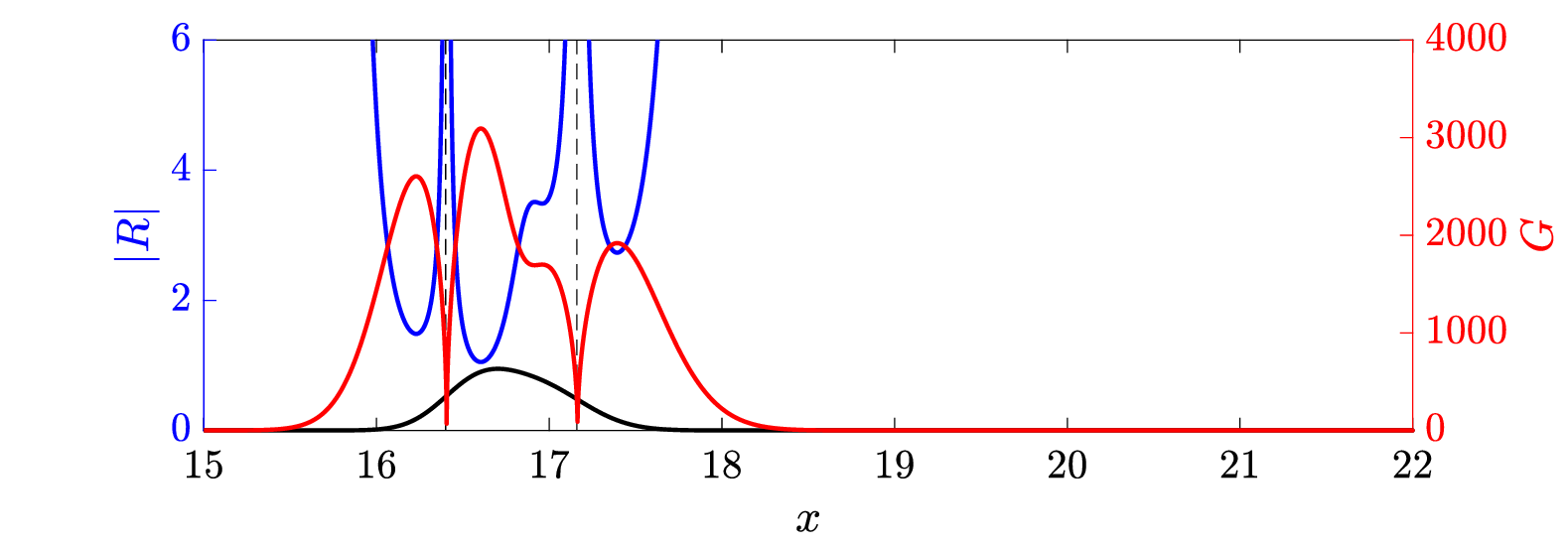}
\includegraphics[width=1\textwidth]{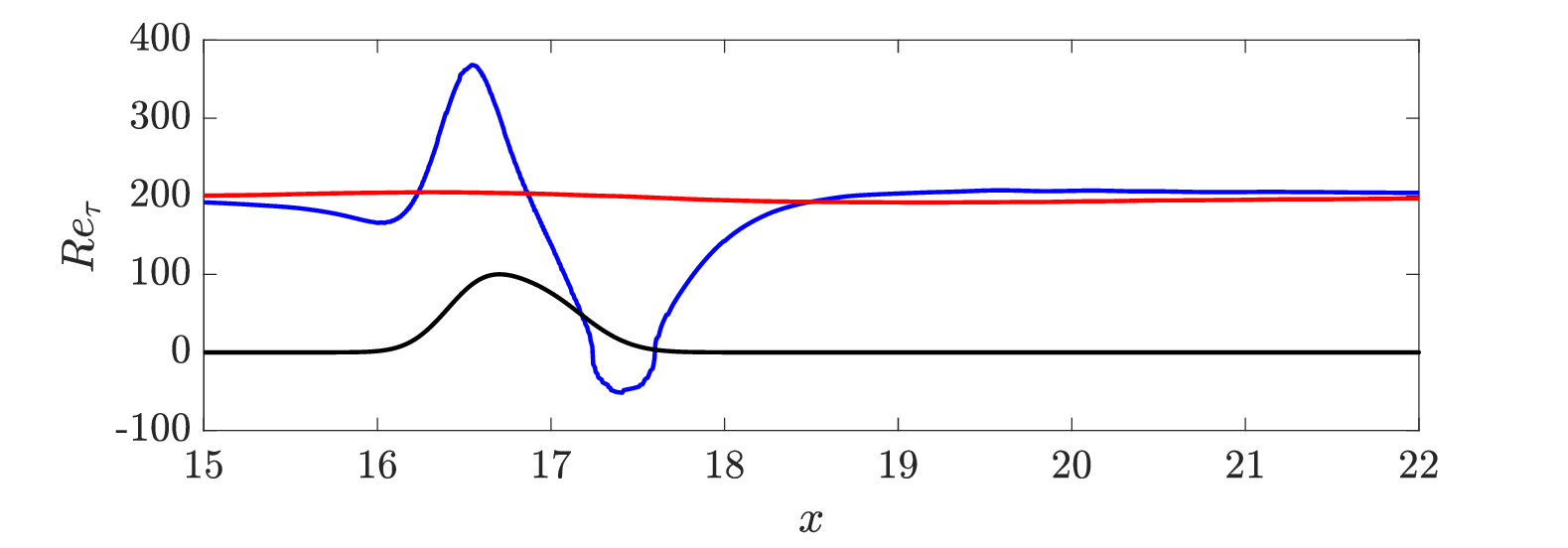}
\caption{Top: streamwise evolution of the (absolute value of the) curvature radius $R$ (blue line), with dashed lines indicating vertical asymptotes; the red line shows for the streamwise evolution of the G\"ortler number, defined in the text. Bottom: streamwise evolution of $Re_\tau$ on the bottom (blue) and top (red) walls. In both plots, the bump is drawn with arbitrary vertical scale.}
\label{fig:radius-ret}
\end{figure}
We start by labeling the main regions of the flow, through a combination of the topology of the mean flow and of the curvature of the wall.

The top panel of figure \ref{fig:radius-ret} shows the evolution along the streamwise direction of the (absolute value of the) local curvature radius $R$ of the wall, computed analytically from Eq.\eqref{eq:bump}. 
The two vertical asymptotes, corresponding to infinite curvature at inflection points in the profile of the bump, indicate that curvature changes sign twice: it is concave for $x < 16.37 $ and $x > 17.16$, and convex for $ 16.37 < x < 17.16$. As mentioned in \S\ref{sec:introduction}, a change of curvature may originate internal layers near the wall.
The top panel also plots the streamwise evolution of the G\"ortler number $G \equiv (U_b h / \nu)\sqrt{h/R}$. 

The bottom panel of figure \ref{fig:radius-ret} plots $Re_\tau$ over the two channel walls. To account for the negative wall shear stress $\tau_w$ in the recirculation zone, here we define it as $Re_\tau = \text{sign} \left( \tau_w \right) \sqrt{ |\tau_w|/\rho } h / \nu$. This plot can be used to convert lengths from outer to local viscous units, and vice versa.
Owing to the minimal blockage, the effect of the bump on the top flat wall is minimal. On the bottom wall, instead, $Re_\tau$ first decreases slightly before the bump, reaching a local (positive) minimum at $x=16.1$. 
Later downstream, $Re_\tau$ quickly rises to a positive maximum of $Re_\tau=368$ ($\approx 1.8$ times the inlet value) at $x=16.55$, slightly before the bump tip. Despite the small height of the bump, a small recirculation bubble is observed after the bump tip; its extension is quantified by the region with $Re_\tau<0$, i.e. $17.24 \le x \le 17.6$. The minimum $Re_\tau=-52$ is found at $x=17.44$. Later on, $Re_\tau$ progressively increases again, and eventually relaxes to the inlet value.

\cite{baskaran-smits-joubert-1987} used the wall-curvature perturbation parameter $\Delta \kappa^+ = (1/R_2-1/R_1) \nu / u_{\tau, 1} $ to quantitatively estimate the influence of a curvature discontinuity across two points on the flow, and observed that an internal layer after the discontinuity is formed whenever $\Delta \kappa^+$ exceeds a threshold value of $0.373 \cdot 10^{-4}$ \citep[see also][]{webster-degraaff-eaton-1996-1}.
This definition of $\Delta \kappa^+$ does not immediately apply here, as our wall profile does not have curvature discontinuities. Nonetheless, we compute the perturbation parameter using the two local minima of $R$ across the curvature changes, which are located at $x \approx 16.23$ and $ x \approx 16.60$ over the upstream bump side, and at $x \approx 16.96$ and $ x \approx 17.39$ over the downstream side. By doing this, we obtain $\Delta \kappa^+=0.090$ for the upstream concave-convex curvature change, and $\Delta \kappa^+=0.025$ for the downstream convex-concave change. 
This suggests that the curvature change is not negligible, and that the upstream change is expected to generate a stronger perturbation on the flow than the downstream one (this will be in fact confirmed later on with the AGKE analysis). 
\begin{figure}
\centering
\includegraphics[trim=0 300 0 150, clip, width=1\textwidth]{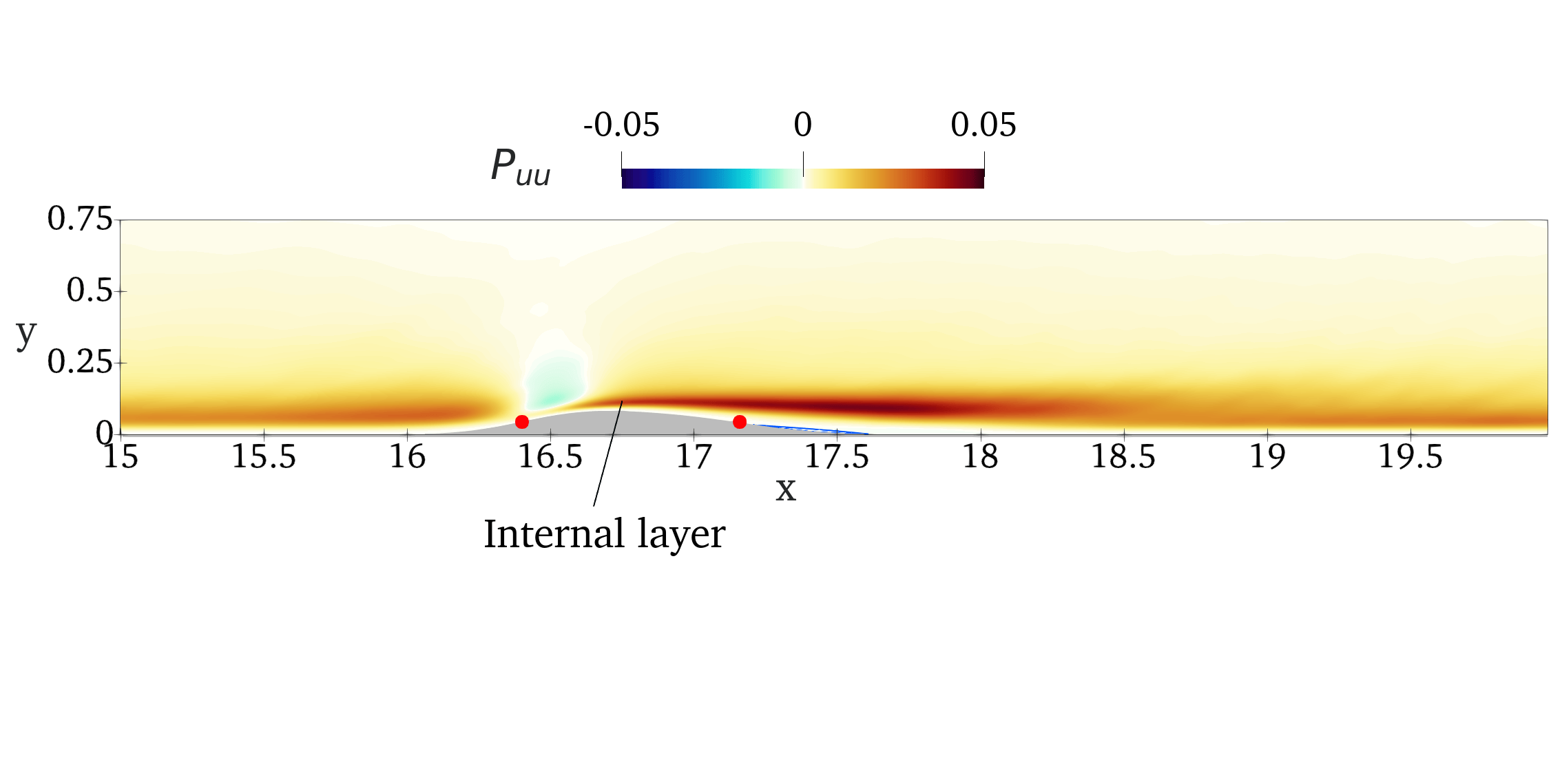}
\caption{Production of the streamwise Reynolds stresses $\aver{uu}$. The region of large production that appears near the upstream curvature change is the trace of a new internal layer \citep{wu-squires-1998}. Red circles indicate the two inflection points.}
\label{fig:prod_uu}
\end{figure}
The internal layer that arises from the upstream curvature change is visible in figure \ref{fig:prod_uu}, where the production term of the single-point streamwise Reynolds stresses $\aver{uu}$, i.e. $P_{uu} = -\aver{uu} \partial U/\partial x - \aver{uv} \partial U/\partial y$ is plotted. Indeed, near the upstream curvature change a new (intense) region of positive production appears and develops downstream; see also figure 12 of \cite{banchetti-luchini-quadrio-2020}, where the production of the complete turbulent kinetic energy is plotted.

\begin{figure}
\centering
\includegraphics[trim=0 330 0 150, clip, width=1\textwidth]{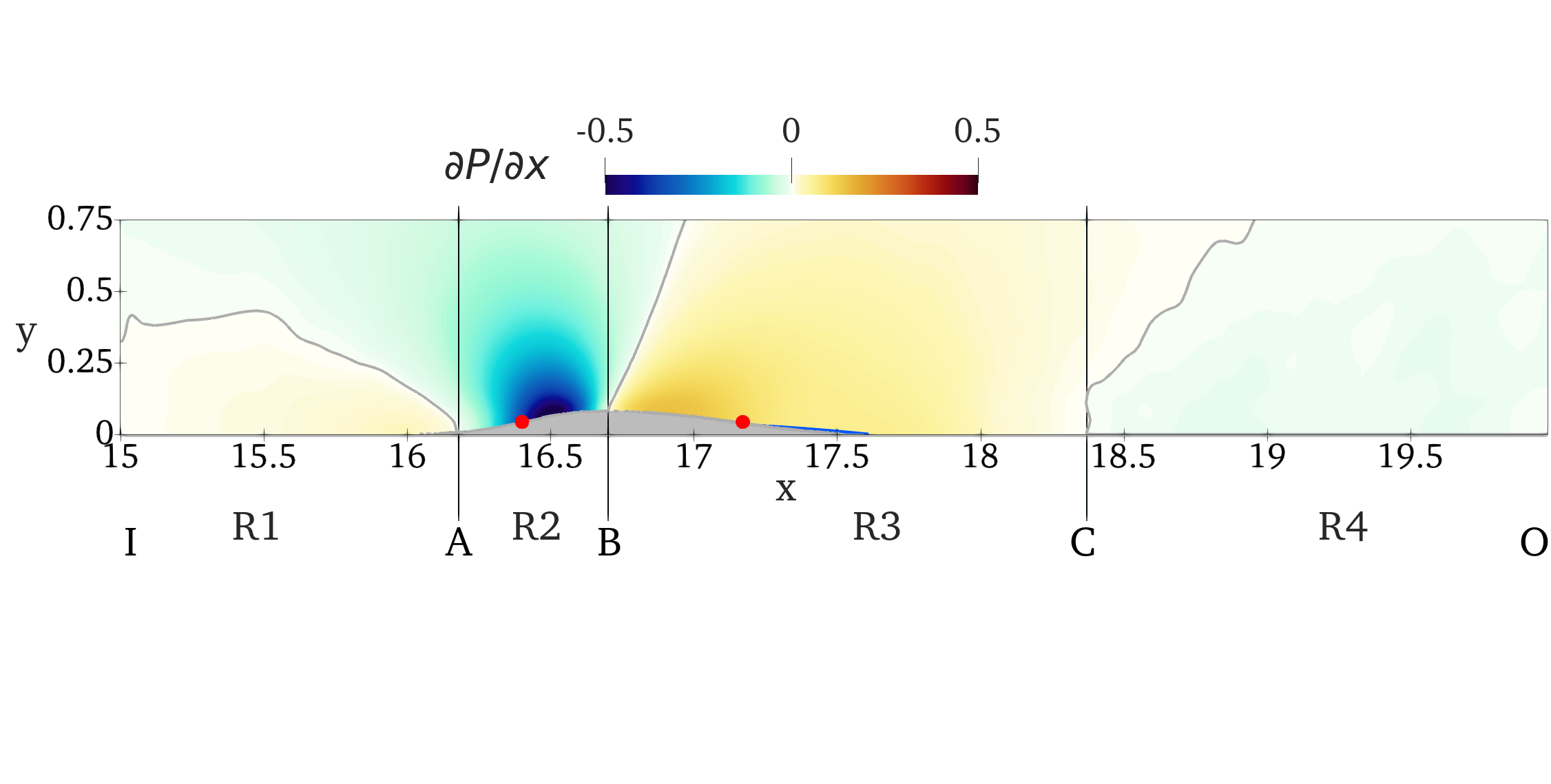}
\caption{Mean streamwise pressure gradient $\partial P / \partial x$. The contour line is for $\partial P/\partial x=0$. The barely visible reverse flow region is shown via the $U=0$ line in blue. Red circles indicate the two inflection points.}
\label{fig:gradp}
\end{figure}
Figure \ref{fig:gradp} plots the mean pressure gradient $\partial P / \partial x$; the letters A, B and C label where $\partial P / \partial x$ at the bottom wall changes sign, while I and O are the inlet and outlet sections respectively. Depending on the sign of the pressure gradient at the wall, the flow over a generic isolated bump can be divided into four regions, labelled in the figure as R1, R2, R3 and R4. R1 goes from the inlet I to A ($x=16.17$), and is characterised by a slightly adverse pressure gradient $\partial P / \partial x >0$ in the near-wall region. R2 goes from A to B ($x=16.71$): here the mean flow accelerates, as the cross-section is diminishing and the pressure gradient is negative (favourable). In R3, extending from B to C ($x=18.37$), the pressure gradient is positive (adverse), and the flow decelerates. Eventually, in R4, which extends from C to the outlet O, the pressure gradient becomes favourable again. The upstream concave-convex change of curvature occurs in R2, while the downstream convex-concave one occurs in R3. When the flow separates, the recirculating bubble is in R3.
\section{The upstream side of the bump: regions R1 and R2}
\label{sec:before}

\subsection{Turbulent structures}
\label{sec:structures-before}

\begin{figure}
\centering
\includegraphics[width=0.95\textwidth]{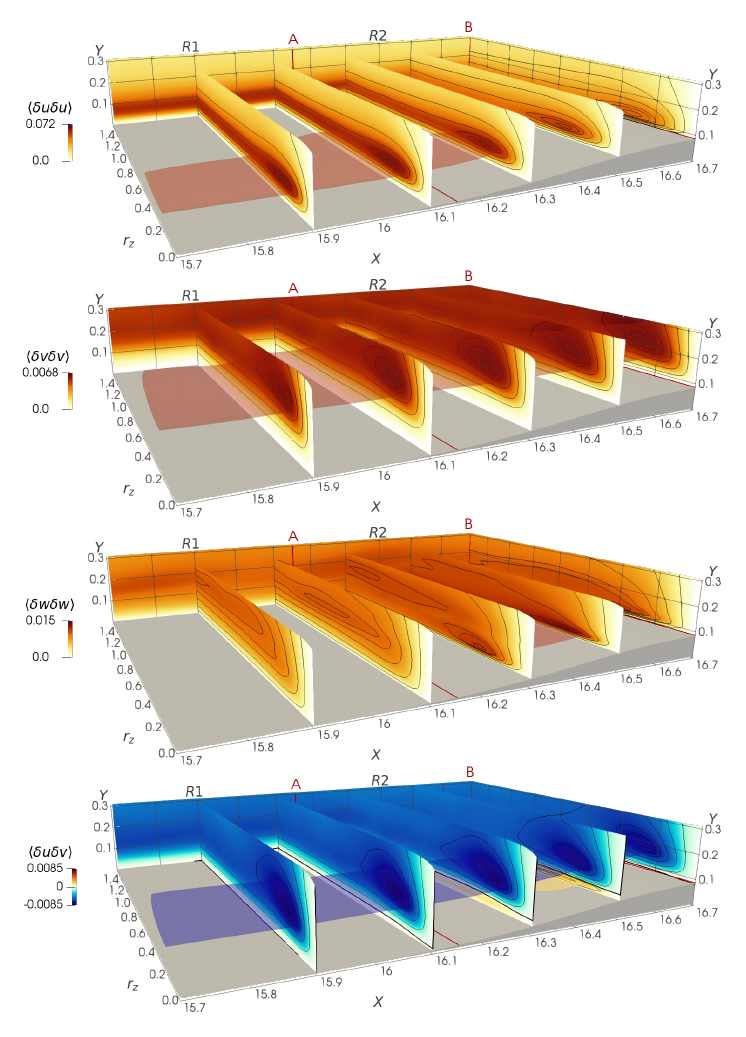}
\caption{Structure functions $\aver{\delta u \delta u}$, $\aver{\delta v \delta v}$, $\aver{\delta w \delta w}$ and $\aver{\delta u \delta v}$ (from top to bottom). Contour lines are drawn at 99\%, 95\%, 90\%, 75\%, 50\%, 20\% of the maximum in each plane, and the isosurfaces are computed for a value of 85\% of the maximum in the volume. The thick black line in the last panel is the zero contour level.}
\label{fig:ener_before}
\end{figure}

\begin{figure}
\centering
\includegraphics[width=0.8\textwidth]{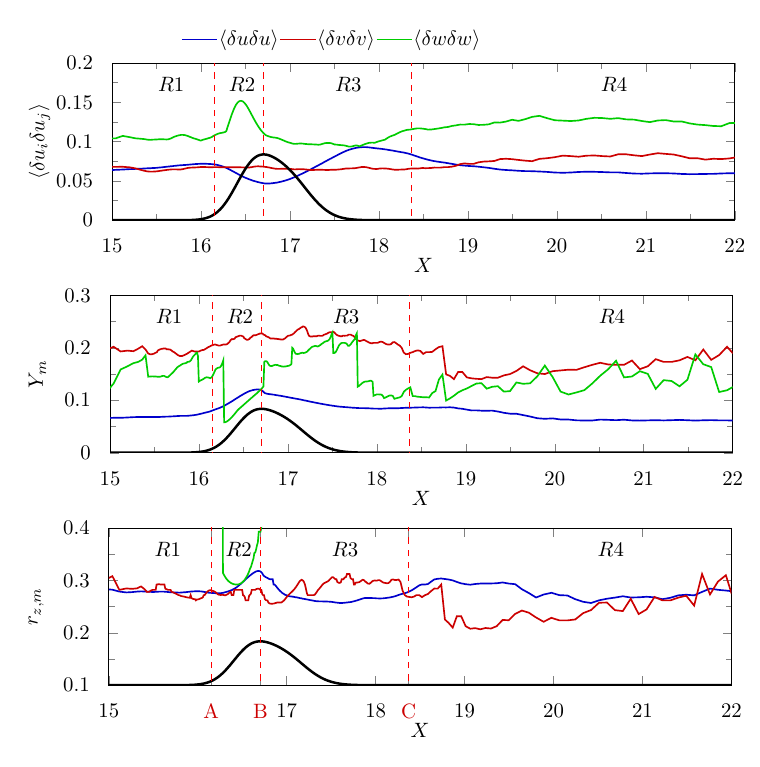}
\caption{Streamwise evolution of the intensity and the $(Y,r_z)$ location of the maxima of the diagonal components of the structure function tensor. The intensity of $\aver{\delta v \delta v}$ and $\aver{\delta w \delta w}$ in the top panel is multiplied by a factor of 10 for visualisation. The black line shows the bump (arbitrary vertical scale). Regions identified by the dashed lines are defined in Figure \ref{fig:gradp}.}
\label{fig:max}
\end{figure}

The turbulent structures populating the flow upstream of the bump tip are observed via the four non-zero components of the structure function tensor, plotted in figure \ref{fig:ener_before} in the $(X,r_z,Y)$ space with $r_x=r_y=0$. 
Figure \ref{fig:max} additionally shows the streamwise evolution of the intensity of local maxima of the diagonal components, $\aver{\delta u_i \delta u_j}_m$ with $i=j$, and of their wall-normal position $Y_m$ and spanwise scale $r_{z,m}$. 
In fact, in this $r_x=r_y=0$ space a positive/negative peak of the structure function $\aver{\delta u_i \delta u_j}$ corresponds to a negative/positive peak of the correlation function $R_{ij}$ (see equation \ref{eq:dudu-corr}), and provides structural information about the flow \citep{jimenez-2018}. 

In the plane channel flow, streaks of high and low streamwise velocity (HSS and LSS) and quasi-streamwise vortices (QSVs) are at the core of the near-wall turbulence cycle. For $r_x=r_y=0$, HSS/LSS and QSVs produce peaks of $\aver{\delta u \delta u}$ and $\aver{\delta v \delta v}$ respectively; the corresponding $r_z$ and $Y$ indicate their characteristic spanwise spacing/size, and wall-normal distance. The streaks, indeed, induce negative $R_{uu}$ at their characteristic $r_z$ spacing, and QSVs induce negative $R_{vv}$ at their lateral sides. (We refer the reader to figure 7 of \cite{gatti-etal-2020}, where the AGKE have been computed from a velocity field induced by the ensemble-averaved quasi-streamwise vortex.)
This is what happens in the upstream periodic portion of the computational domain before the inlet I, where peaks of $\aver{\delta u \delta u}$ and $\aver{\delta v \delta v}$ are found at $(r_z,Y)=(0.28,0.07)$ and $(r_z,Y)=(0.3,0.21)$, or $(r_z^+,Y^+)=(55,14)$ and $(r_z^+,Y^+)=(59,42)$. The classical spacing $r_z^+ \approx 100$ between LSS \citep{kim-moin-moser-1987, robinson-1991b} is twice the $r_z^+\approx 50$ separation of maximum negative $R_{uu}$ correlation between the low- and high-speed streaks. In this region, $\aver{\delta u \delta u}$ is larger than $\aver{\delta v \delta v}$ and $\aver{\delta w \delta w}$, indicating that streamwise velocity fluctuations are dominant over the cross-stream ones.

In region R1, the mild adverse pressure gradient determined by the bump curvature intensifies the streamwise fluctuations, with $\aver{\delta u \delta u}_m$ increasing by 15\% from $X=15.5$ to $X=16$. Wall-normal fluctuations are affected too, but only marginally, with $\aver{\delta v \delta v}_m$ increasing by 3\%. Such intensification of turbulent fluctuations confirms the result of \cite{wu-squires-1998}, who found a local peak of the turbulent kinetic energy $k$ just upstream of the bump.

In region R2, where the pressure gradient becomes favourable again, the streaks weaken as the magnitude of $\aver{\delta u \delta u}$ decreases, and their spanwise spacing increases by 14\% from $r_{z,m}=0.28$ at $X=16.37$ (at zero curvature) to $r_{z,m}=0.32$ at $X=16.7$ (at the bump tip).
This differs from the results of \cite{marquillie-laval-dolganov-2008}, who observed for a higher bump a slight decrease of the spanwise spacing of the streaks, measured in outer units. 
If the spacing is expressed in viscous units computed with the local $u_\tau$, its increase becomes even more evident: $r_z^+=81$ at $X=16.37$ goes up to $r_z^+=112$ at $X=16.55$, the position of maximum $Re_\tau$, to decrease again to $r_z^+=97$ at the tip. 
The lack of viscous scaling indicates that the streaks do not immediately respond to the changing wall friction \citep{marquillie-laval-dolganov-2008}. Moreover, $\Delta Y_m$ decreases along the bump, meaning that the streaks are pushed towards the wall. 

Unlike the streaks, QSVs are marginally affected by the bump, as seen from the map of $\aver{\delta v \delta v}$: only a small increase of $\aver{\delta v \delta v}_m$ and a decrease of $r_{z,m}$ close to the bump tip can be noticed. 
This difference is explained by the different typical wall distance at which the two types of structures reside. The bump height $h_b=0.0837$, or $h_b^+=25$ ($h_b^+=17$ when computed with the friction velocity of the planar channel), is in fact comparable with the average position of the streaks, but lower than that of the QSVs, which exist at $Y \approx 0.2$ or $Y^+ \approx 40$ in the planar channel region \citep{jeong-etal-1997}. 

\begin{figure}
\centering
\includegraphics[width=1\textwidth]{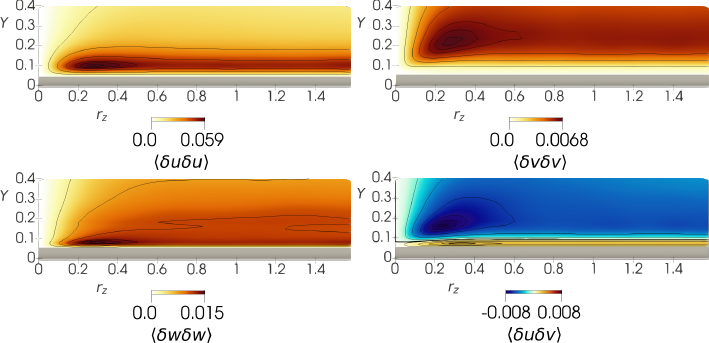}
\caption{Structure functions on a plane at $X=16.44$. Contour lines as in figure \ref{fig:ener_before}.}
\label{fig:uphill_1644}
\end{figure}

In R2, the presence of the bump leads to a new statistical feature, related to a type of turbulent structure which does not exist over a plane wall. 
Figure \ref{fig:uphill_1644} shows a cross-section at $X=16.44$, i.e. near the end of the concave portion of the bump, and highlights how structures of spanwise velocity are generated close to the bump surface. 
Here (and in the whole range $16.28 \le X \le 16.7$) the map of $\aver{\delta w \delta w}$ shows a distinct near-wall positive peak at $r_{z,m} \approx 0.3-0.4$ and $\Delta Y_m \approx 0.03-0.04$, indicating structures with negatively correlated spanwise velocity for $r_z \neq 0$. The peak cannot derive from the QSVs, which produce a negative $R_{ww}$ only when $r_y \neq 0$ \citep{schoppa-hussain-2002, gatti-etal-2020}. 
Instead, the observed maximum is consistent with the incipient development of Taylor--G\"ortler vortices (TGVs). They are pairs of near-wall counter-rotating vortices, which originate from an inviscid instability mechanism over concave surfaces, because of the imbalance between the centrifugal force and the radial pressure gradient \citep{gortler-1941, floryan-1991, saric-1994, xu-liu-wu-2020}. In laminar boundary layers, TGVs are known to advect upwards/downwards low/high momentum, and to induce streamwise velocity fluctuations larger than spanwise ones \citep{xu-etal-2017,xu-liu-wu-2020}. However, the contribution of the TGVs to $\aver{\delta u \delta u}$ cannot be distinguished from that of the streaks as the latter structures induce much larger fluctuations at a similar spanwise scale (see figure \ref{fig:uphill_1644}).

The presence of these vortices is usually assessed using the G\"ortler number $G$, which is very large in the concave portion of the bump: using $h$ and $U_b$ as reference length and velocity scales, the G\"ortler number is approximately $G = Re_b (h/R)^{1/2} \approx 3000$ at $x=16.6$ where $R$ is minimum (see the top panel of figure \ref{fig:radius-ret}). According to the linear stability theory, in a laminar boundary layer the first onset of this instability is experienced at $G=(U_\infty \delta/\nu)(\delta/R)^{1/2}=0.4638$, where $\delta=(\nu x_0/U_\infty)^{1/2}$ is the local boundary layer thickness at $x=x_0$ \citep{floryan-saric-1982}. 
Note that in the turbulent regime, if present, TGVs are immersed in a random background that makes their identification more difficult. Although their presence causes a near-wall peak in the profile of the single-point spanwise Reynolds stress $\aver{ww}$ (result not shown), the absence of the scale information prevents their certain identification. Hence, the scale-space information of double-point statistics is crucial for their detection.



The peak of $\aver{\delta w \delta w}$ moves towards larger $r_z$ along the bump, while its wall distance increases but only slightly; it is located at $(r_{z,m},\Delta Y_m)=(0.29,0.03)$ for $X=16.37$, and at $(r_{z,m},\Delta Y_m)=(0.39,0.04)$ for $X=16.7$. 
This is again consistent with the evolution of the TGVs, which progressively move apart in the spanwise direction and increase in size \citep{xu-liu-wu-2020}.

As the streamwise evolution of the magnitude of $\aver{\delta w \delta w}_m$ confirms, these structures are sustained only over the concave portion of the bump, whereas the convex wall inhibits their growth \citep{benmalek-saric-1994}: here the TGVs are progressively damped by the redistribution via pressure--strain (see \S\ref{sec:pstr-before}).
When studying the non linear development of TGVs over walls with variable curvature, indeed, \cite{benmalek-saric-1994} found that a convex wall reduces their growth more than a flat wall \citep{peerhossaini-wesfreid-1988}, and eliminates the three-dimensional structures of streamwise velocity.

A detailed description of the sustaining mechanism of TGVs is provided later in \S\ref{sec:pstr-before}, where the scale-space energy redistribution is addressed.

\subsection{Production of turbulent kinetic energy} 
\label{sec:prod-before}

\begin{figure}
\centering
\includegraphics[width=1\textwidth]{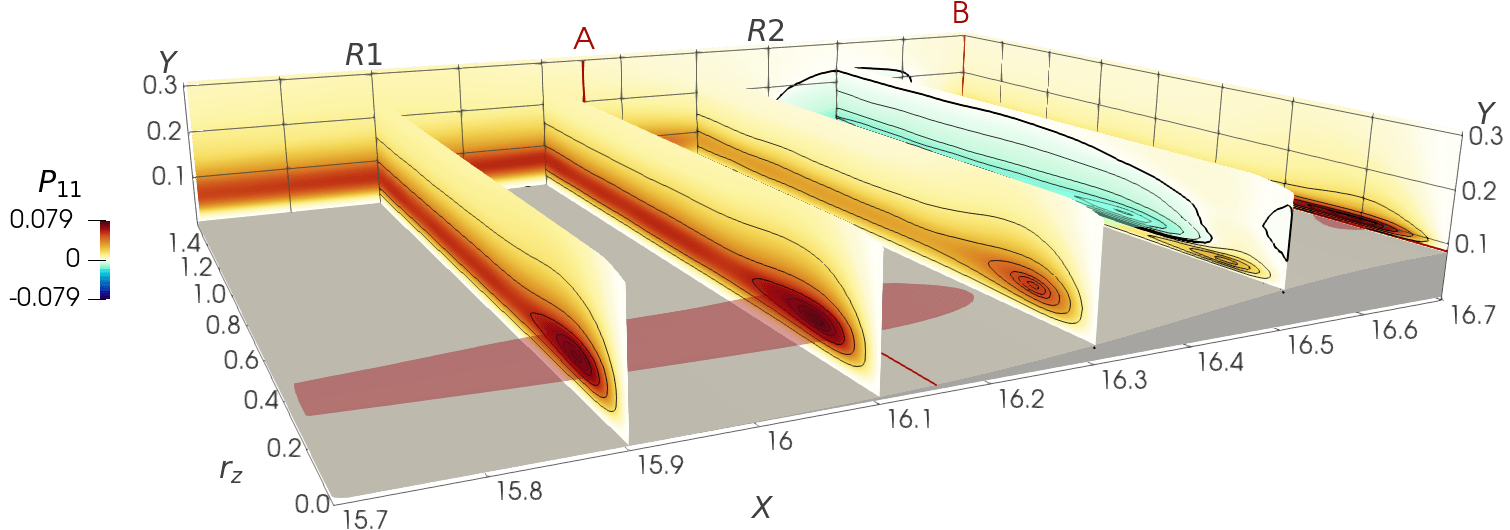}
\includegraphics[width=1\textwidth]{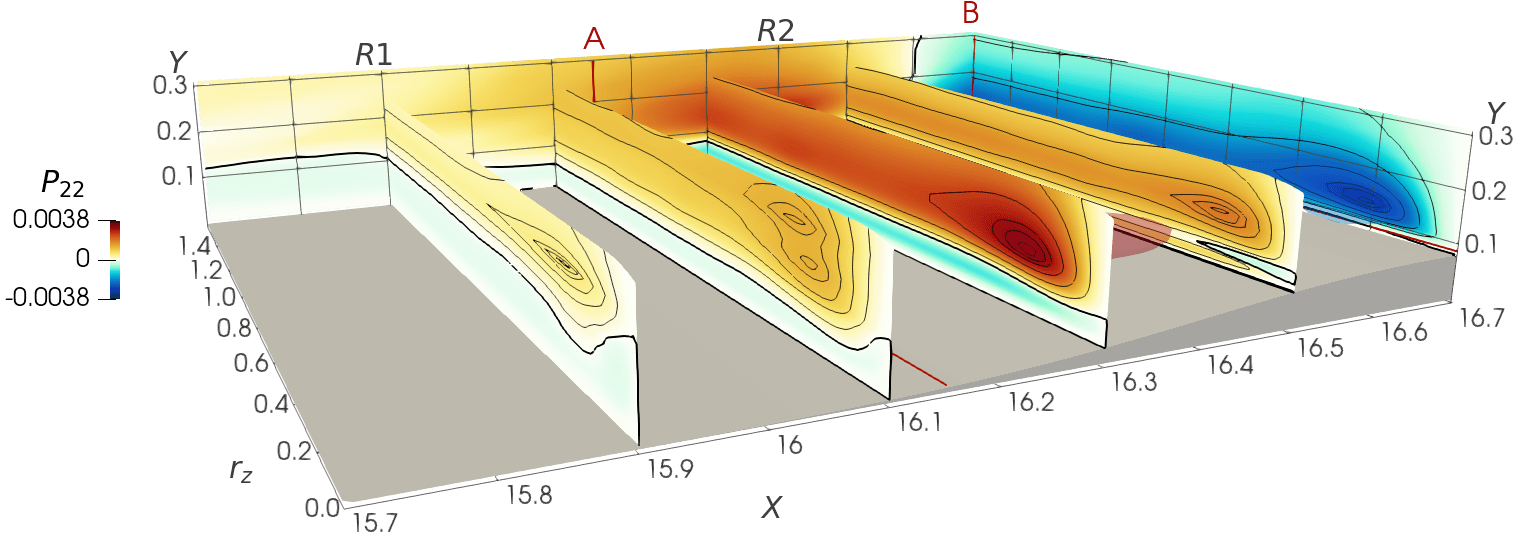}
\caption{Production terms $P_{11}$ (top) and $P_{22}$ (bottom). See caption of figure \ref{fig:ener_before} for contour lines and isosurfaces.}
\label{fig:prod_before}
\end{figure}

The components of the turbulent kinetic energy production tensor are shown in figure \ref{fig:prod_before}. Due to the spanwise homogeneity, $W$ and its derivatives are zero, so $P_{33}=0$ everywhere. 
%

In R1, production resembles that in a plane channel flow: $P_{11}$ is positive at all scales and positions and it is one order of magnitude larger than $P_{22}$. Hence, energy is transferred from the mean field mainly towards the streamwise fluctuating field, feeding the streaks. In a plane channel, indeed, QSVs (responsible for cross-stream fluctuations) are known to be energised by redistribution only \citep{gatti-etal-2020}.

Over a curved wall, production is known to differ from the plane wall \citep{smits-wood-1985}. In R2 the amount of streamwise energy drained from the mean flow to feed the streaks decreases, in agreement with their loss of coherency observed above in \S\ref{sec:structures-before}. 
The maximum production $P_{11,m}$ decreases along the streamwise coordinate; its evolution resembles that of the scale shear stresses $\aver{\delta u \delta v}$ ($- 2 \aver{\delta u \delta v} \partial U/ \partial y$ is the dominant contribution here), i.e. moves towards smaller $r_z$ and $\Delta Y$. 

Careful inspection of figure \ref{fig:prod_before} reveals a second region of positive $P_{11}$ close to the wall, just after the concave-convex curvature change; see the peak of $P_{11}$ at $(X,r_z, \Delta Y) \approx (16.5, 0.14, 0.016)$ (see also figure \ref{fig:prod_uu}). 
This production sustains the streamwise fluctuations in the upstream internal layer, and is responsible for the positive production of turbulent kinetic energy downstream of the bump, shown for example in figure 12 of \cite{banchetti-luchini-quadrio-2020}. Above this layer, $P_{11}$ is negative for $16.4 \le X \le 16.6$ and $r_z>0.4$, due to the positive $\partial U / \partial x$ (not shown), with a local minimum that identifies the spanwise scale $r_z \approx 0.42$. 
For these scales and positions, the interaction between the near-wall cycle and the changing mean velocity field leads to a sink for the streamwise fluctuations; here the turbulent activity is reduced as the mean flow accelerates. This is in agreement with the relaminarisation of the boundary layer over a bump observed by other authors in the same region \citep[see for example][]{webster-degraaff-eaton-1996-2}.



The term $P_{22}$, i.e. the production of $\aver{\delta v \delta v}$, is one order of magnitude smaller than $P_{11}$ at all scales and positions; this is to be expected, owing to the small height of the bump considered in this work.
In fact, the velocity derivatives $\partial V / \partial x$ and $\partial V / \partial y$, and thus $P_{22}$, are determined by the geometry of the bump, as already shown for example by \cite{bradshaw-1973}, who used $\partial V/\partial x$ as a proxy for the curvature of the wall. 
In R1 and in the part of R2 upstream of the inflection point, the concave wall induces $P_{22}>0$ for $Y \gtrapprox 0.1$, and $P_{22}<0$ close to the wall. When $P_{22}>0$, the mean flow feeds not only the streamwise velocity streaks (as in the canonical channel flow) but the QSVs too. When $P_{22}<0$, instead, energy is drained from the $v$ fluctuations to feed the mean field.
%
After the concave-convex change of curvature, the region with $P_{22}<0$ shrinks and then grows again over the bump tip, where it extends for all $r_z$ and $\Delta Y \gtrapprox 0.02$, with a clear peak associated with the QSVs at $(r_z,\Delta Y) \approx (0.22,0.05)$, or $Y \approx 0.13$.
This negative $P_{22}$ resembles what was observed slightly upstream for $P_{11}$, indicating a net sink for both the streamwise and vertical velocity fluctuations, consistently with the above-mentioned tendency of the flow to laminarise near the bump tip. 

In the internal layer that originates at the inflection point (see \S\ \ref{sec:flow-topology} and figure \ref{fig:prod_uu}), a small region of positive $P_{22}$ arises close to the surface, and shrinks at the bump tip (it then grows again in regions R3 and R4; see later \S\ref{sec:prod-after}). In the internal layer, therefore, both streamwise and vertical fluctuations are sustained by positive production. However, the $P_{22}>0$ region extends for a lower $\Delta Y$ range and is weaker compared to $P_{11}$: overall, the production of vertical fluctuations is less affected by the internal layer. This is again consistent with the results of \cite{wu-squires-1998}.

\subsection{Redistribution of turbulent kinetic energy} 
\label{sec:pstr-before}
\begin{figure}
\centering
\includegraphics[width = \textwidth]{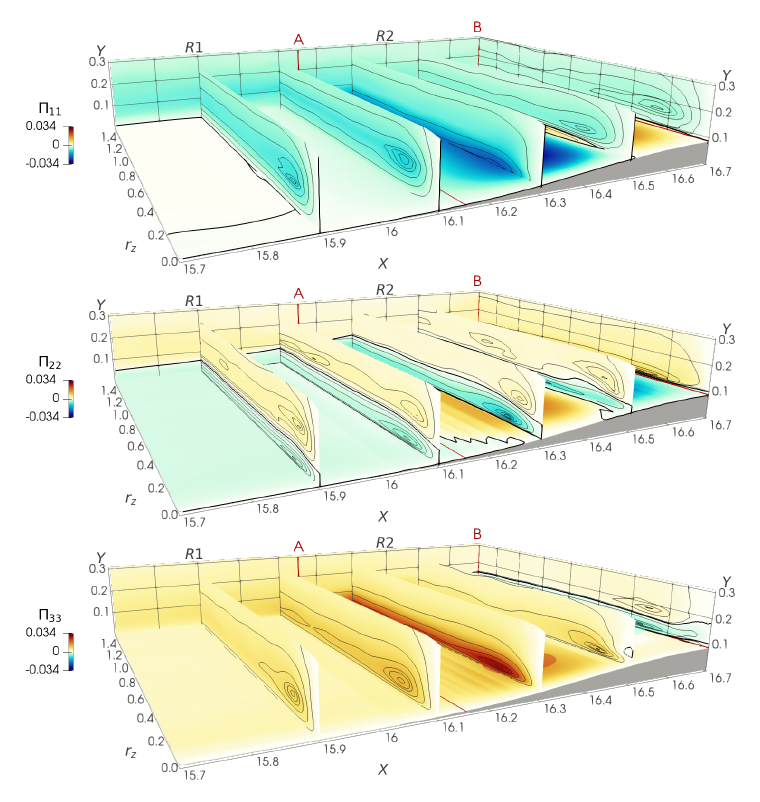}
\caption{Pressure strain terms $\Pi_{11}$ (top), $\Pi_{22}$ (middle) and $\Pi_{33}$ (bottom). See caption of figure \ref{fig:ener_before} for contour lines and isosurfaces.}
\label{fig:pstr_before}
\end{figure}
Lastly, we consider the energy redistribution with figure \ref{fig:pstr_before}, which shows the pressure--strain terms for the three normal components over the upstream half of the bump.

In the plane channel and in R1, two redistribution mechanisms take place \citep{gatti-etal-2020}. For $\Delta Y \gtrapprox 0.1$, part of the energy drained by the mean flow to feed the streamwise fluctuations is redistributed towards the cross-stream components to feed the QSVs. This is seen here in the $X=15.9$ plane, via the negative peak of $\Pi_{11}$ at $(r_z,\Delta Y) \approx (0.23,0.09)$ and the positive peaks of $\Pi_{22}$ and $\Pi_{33}$ at $(r_z,\Delta Y)=(0.17,0.13)$ and $(r_z,\Delta Y)=(0.23,0.07)$. Close to the wall, instead, the vertical fluctuations are redistributed towards the in-plane ones owing to the so-called splatting \citep{mansour-kim-moin-1988}; here $\Pi_{11}>0$, $\Pi_{22}<0$ and $\Pi_{33}>0$ (see the colour maps in the $Y \rightarrow 0$ plane in figure \ref{fig:pstr_before}).

In R2, an additional and intense redistribution takes place close to the wall. In the concave part, streamwise and vertical contributions to the fluctuating energy are redistributed towards the spanwise one: this is shown at $X=16.3$ by the negative peaks of $\Pi_{11}$ and $\Pi_{22}$ very near the wall at $r_z \approx 0.3$ and $r_z \approx 0.2$ respectively, accompanied by the positive peak of $\Pi_{33}$ at $(r_z,\Delta Y) \approx (0.24,0.02)$. 
The scale $r_z \approx 0.2-0.3$ and the wall distance $\Delta Y \approx 0.02$ are consistent with the position of local maximum of $\aver{\delta w \delta w}$ discussed above in \S\ref{sec:structures-before} and associated with the TGVs.
Hence, the generation of TGVs in the concave part of the bump is accompanied, in the vicinity of the wall, by an intense reorientation of both streamwise and vertical fluctuations into spanwise ones.
In particular, the main contribution derives from the streamwise fluctuations, since $|\Pi_{11}|>|\Pi_{22}|$ there. This is consistent with the decrease of $\aver{\delta u \delta u}_m$ and the increase of $\aver{\delta w \delta w}_m$ observed in figure \ref{fig:max} at these streamwise coordinates.

After the inflection point, the redistribution changes nature: near the bump tip streamwise and spanwise fluctuations are reoriented into vertical ones. Indeed, $\Pi_{11}<0$, $\Pi_{22}>0$ and $\Pi_{33}<0$ for $0.1 \le Y \le 0.15$: at $X=16.6$ for example, their peak is found at $(r_z,\Delta Y) \approx (0.3,0.02-0.08)$.
This is consistent with the observation that here $\aver{\delta w \delta w}_m$ decreases with $X$, while $\aver{\delta v \delta v}_m$ increases (see figure \ref{fig:ener_before}). Therefore, along the convex portion of the bump the TGVs progressively vanish: they are not sustained anymore, and their energy feeds mostly the vertical fluctuations.

\section{The downstream side of the bump: regions R3 and R4}
\label{sec:after}

\subsection{Turbulent structures}
\label{sec:structures-after}
\begin{figure}
\centering
\includegraphics[width=1\textwidth]{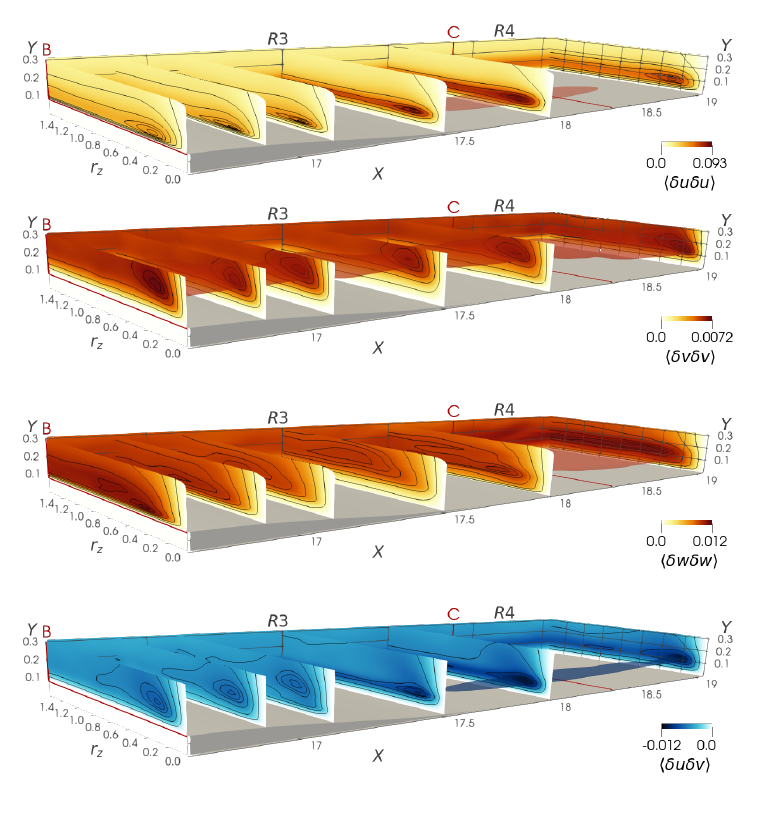}
\caption{Structure functions $\aver{\delta u \delta u}$, $\aver{\delta v \delta v}$, $\aver{\delta w \delta w}$ and $\aver{\delta u \delta v}$ (from top to bottom). Contour lines are drawn at 99\%, 95\%, 90\%, 75\%, 50\%, 20\% of the maximum in each plane, and isosurfaces correspond to 85\% of the maximum in the volume.}
\label{fig:ener_after}
\end{figure}

In analogy with figure \ref{fig:ener_before}, which dealt with the upstream part of the bump, figure \ref{fig:ener_after} plots the components of the structure function tensor for the downstream part, which includes regions R3 and R4.

The upstream internal layer, generated upstream in R2, grows rapidly in R3, owing to the adverse pressure gradient, and influences the spatial organisation of the turbulent fluctuations \citep{webster-degraaff-eaton-1996-1, wu-squires-1998}.
The velocity streaks become more vigorous and tend to decrease their characteristic spanwise spacing. In fact, the maximum $\aver{\delta u \delta u}_m$ increases with $X$, and its spanwise scale $r_{z,m}$ shrinks, especially if quantified in local viscous units (see also figure \ref{fig:max}). 
This should be contrasted with the opposite behaviour observed in R2, where the pressure gradient is favourable.

QSVs are weakly affected in R3 (like in R2): $\aver{\delta v \delta v}_m$ and its scale $r_{z,m}$ and wall distance $Y_m$ remain almost constant with $X$. The map of $\aver{\delta w \delta w}$, moreover, shows that the TGVs generated upstream quickly vanish after the bump tip; the local maximum of $\aver{\delta w \delta w}$ close to the wall at $r_z \approx 0.2$, indeed, disappears for $X \ge 16.9$.


Close to the downstream convex-concave change of curvature at $X \approx 17.1$, a second internal layer is formed, in agreement with previous observations \citep[see for example][]{webster-degraaff-eaton-1996-1}. Unlike the first internal layer, however, this one has a minor effect on the velocity fluctuations, because it develops across a smaller pressure gradient and for a shorter spatial extent. This is consistent with the wall-curvature parameter $\Delta \kappa^+$ being smaller here than upstream (see \S\ref{sec:flow-topology}).
%


Further downstream, in the (small) recirculating region for $16.3 \le X \le 17.6$ the near-wall contribution of $\aver{\delta w \delta w}$ to the turbulent kinetic energy increases. This is due to the impinging flow upon reattachment, which reorients vertical and streamwise fluctuations into spanwise ones \citep[see for example][and section \S\ref{sec:pstr-after}]{chiarini-etal-2022}. 
The recirculation affects streaks and QSVs only marginally. Figures \ref{fig:ener_after} and \ref{fig:max} confirm that there is no downstream evolution of the maxima $\aver{\delta u_i \delta u_j}_m$ for $16.3 \le X \le 17.6$. 
Clearly, this is due to the marginal recirculation created by the tiny bump. \cite{mollicone-etal-2018}, for example, found that a higher bump with a larger recirculation leads to a streamwise and wall-normal stretching of the flow structures.

Lastly, in R4 the flow is attached again, and gradually recovers towards the canonical plane channel flow. 
However, as visible in figure \ref{fig:max}, during the recovery $\aver{\delta u \delta u}$, $\aver{\delta v \delta v}$ and $\aver{\delta w \delta w}$ peak at different streamwise positions, i.e. $X=17.8$, $X=21$ and $X=20$ respectively. 
The energy associated with the streamwise velocity streaks is maximum just after the reattachment, where streaks have been intensified by the adverse pressure gradient in R3. 
The peak position of $\aver{\delta v \delta v}_m$ and $\aver{\delta w \delta w}_m$, instead, are found more downstream. This is consistent with the results of \cite{marquillie-ehrenstein-laval-2011}. They observed that, near the peak of the turbulent kinetic energy $k$ just downstream the bump, which coincides with the peak of its largest contributor $\aver{\delta u \delta u}$, the streaks become unstable and break down into smaller structures, which further downstream evolve into hairpin-like streamwise vortices, so that the reorganisation of the fluctuations in the vertical and spanwise components occurs later than the streamwise one.

\subsection{Production of turbulent kinetic energy} 
\label{sec:prod-after}

\begin{figure}
\centering
\includegraphics[width = 1 \textwidth]{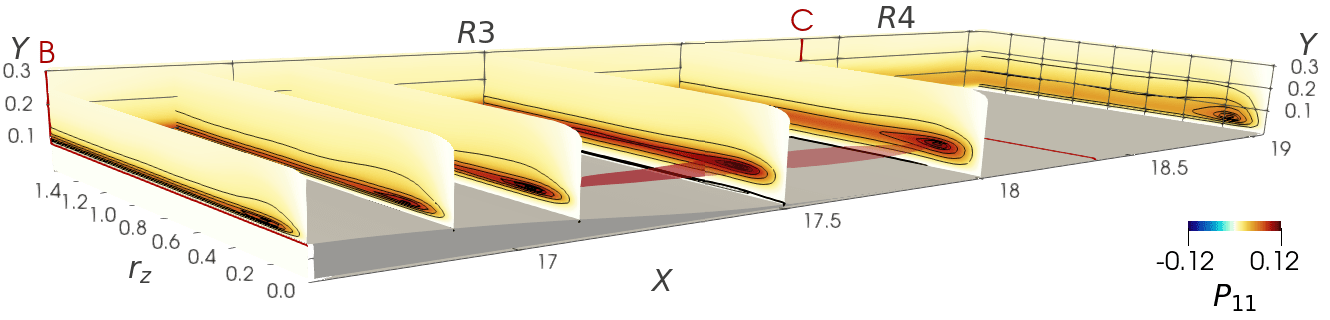}
\includegraphics[width = 1 \textwidth]{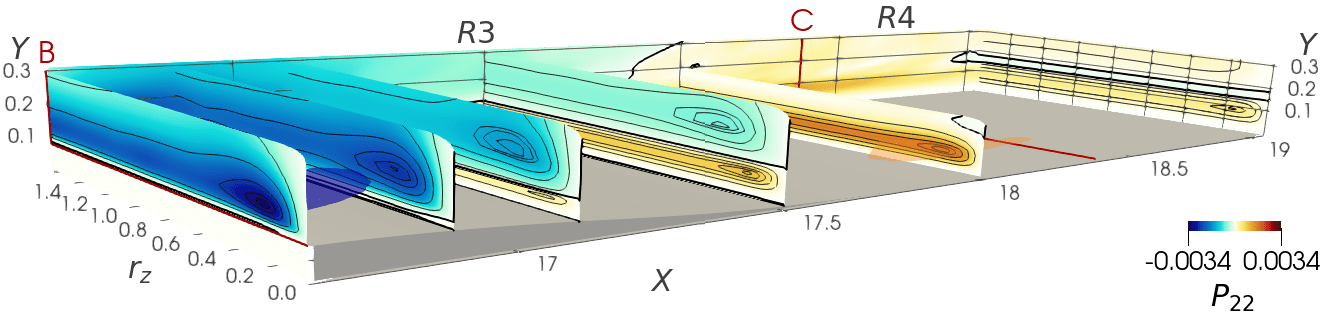}
\caption{Production terms $P_{11}$ (top) and $P_{22}$ (bottom). See caption of figure \ref{fig:ener_after} for contour lines and isosurfaces.}
\label{fig:prod_after}
\end{figure}
\begin{figure}
\centering
\includegraphics[trim={0 200 0 100},clip,width=0.9\textwidth]{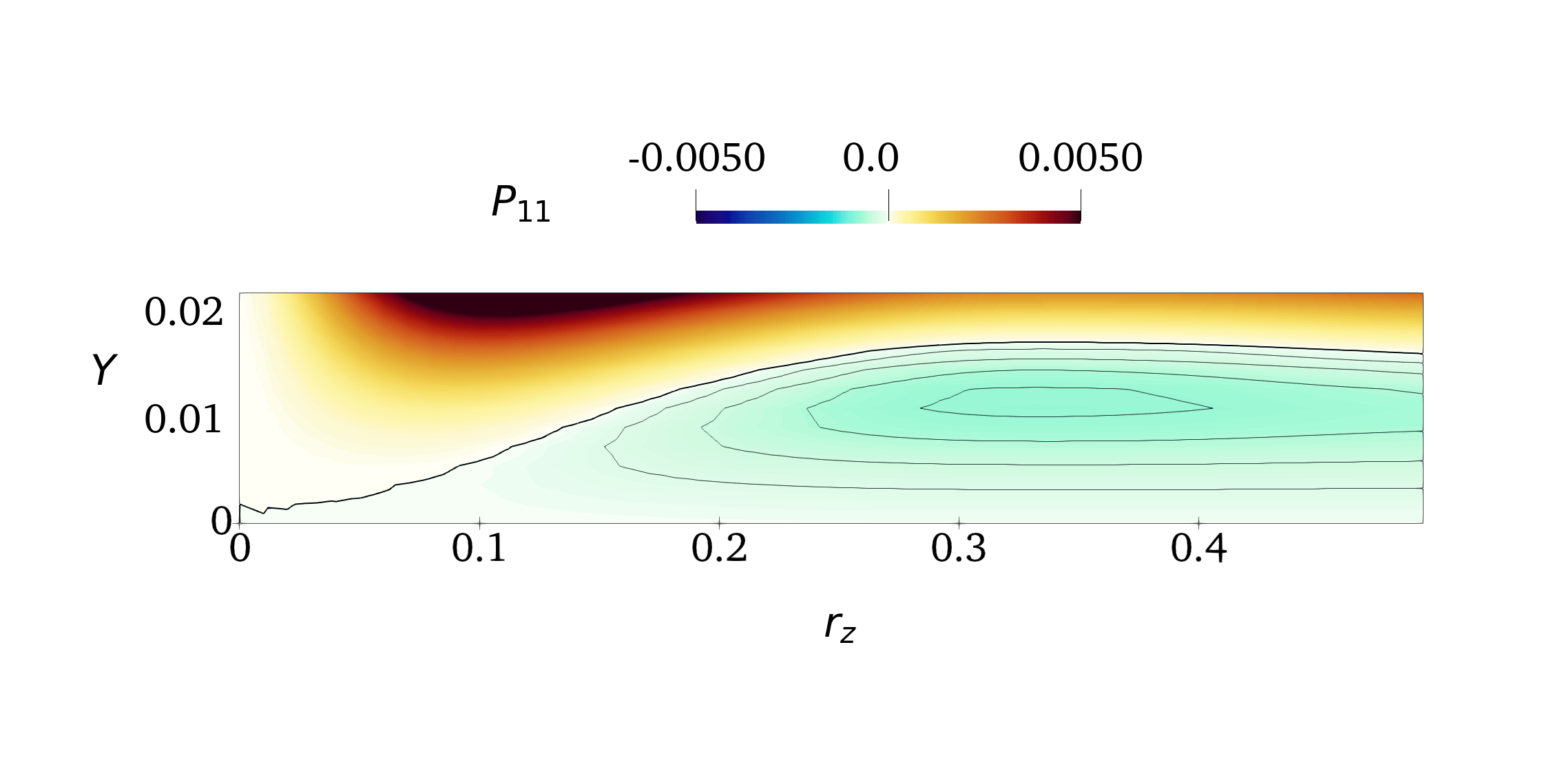 }
\caption{Zoom of the production term $P_{11}$ in the recirculating bubble at $X=17.75$. The black thick line is for $P_{11}=0$. The black thin lines indicate isovalues of $P_{11}$, with an increase of $0.0002$.}
\label{fig:prod11_zoom}
\end{figure}
%

Figure \ref{fig:prod_after} plots the production terms for $\aver{\delta u \delta u}$ and $\aver{\delta v \delta v}$ in regions R3 and R4.
The map of $P_{11}$ shows that streamwise energy is drained from the mean flow to feed the streamwise fluctuating field almost everywhere, except in a thin near-wall region close to the mean flow reattachment point (see figure \ref{fig:prod11_zoom}). 
The scales and wall-normal distances identified by the positive local maxima of $P_{11}$, i.e. $r_z \approx 0.2$ and $\Delta Y \approx 0.02$, indicate that the streamwise velocity fluctuations are again mainly sustained by the interaction of the wall cycle with the mean flow. 
In R3, $P_{11}$ increases with $X$, consistently with the strengthening of the streaks observed above in \S\ref{sec:structures-after}; this is also shown by the downstream evolution of the local maximum of $P_{11}$, which is $P_{11,m}=0.096$ at $X=17$ and increases to $P_{11,m}=0.12$ at $X=17.5$. 
The global maximum of $P_{11}$ is found close to the maximum of $\aver{\delta u \delta u}$, just after the flow reattachment at $(X,r_z,Y) =(17.68,0.2,0.09)$, in agreement with the results of \cite{marquillie-ehrenstein-laval-2011}. 

In the near-wall region close to the flow reattachment point, i.e between $17.45 \le X \le 18$ at $ \Delta Y \le 0.02$, streamwise energy is drained from the fluctuating field to feed the mean flow at all scales: figure \ref{fig:prod_after} shows a very thin region with slightly negative $P_{11}$ (see also the near-wall zoom of figure \ref{fig:prod11_zoom}). This negative region results from the positive $\partial U / \partial x$, which corresponds to an acceleration of the reverse/forward near-wall mean flow in the region downstream/upstream of the reattachment point. 
Interestingly, the same negative production due to the mean flow acceleration has been observed in the separating and reattaching flow past an elongated rectangular cylinder \citep{chiarini-etal-2022}; this suggests that a near-wall region with negative production is a general feature for reattaching flows.

The production term for $\aver{\delta v \delta v}$ is almost two orders of magnitude smaller than $P_{11}$. In R3, $P_{22}$ is negative, with its local minima related to QSVs. 
Hence, QSVs are energised by the mean flow in the upstream side of the bump, but release energy back to the mean flow in the downstream side. 
The negative peak of $P_{22}$ moves away from the wall as $X$ increases.
At the same time, a positive $P_{22}$ appears near the wall, close to the convex-concave change of curvature. This layer with $P_{22}>0$ thickens, and its intensity increases, until in R4 (for $X \gtrapprox 18$) it extends for all $Y$. The maximum intensity of $P_{22}$ occurs at $(X,r_z,Y) \approx (18,0.18,0.07)$. Further downstream, it gradually decreases to become eventually zero as for the plane channel flow.


\subsection{Redistribution of turbulent kinetic energy} 
\label{sec:pstr-after}

\begin{figure}
\centering
\includegraphics[width = 1 \textwidth]{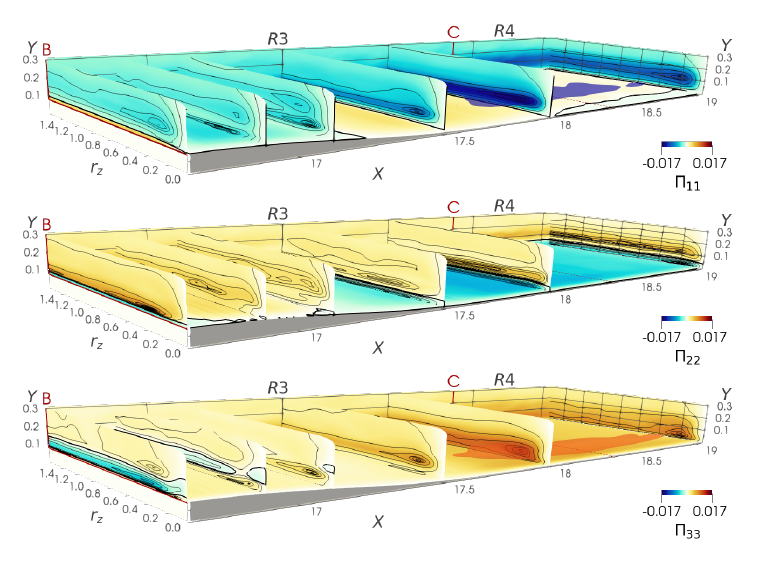}
\caption{Pressure--strain terms $\Pi_{11}$ (top), $\Pi_{22}$ (center) and $\Pi_{33}$ (bottom). See caption of figure \ref{fig:ener_after} for contour lines and isosurfaces.}
\label{fig:pstr_after}
\end{figure}

Figure \ref{fig:pstr_after} plots the pressure--strain terms for the three normal components of the structure function tensor. Four different redistribution mechanisms take place in regions R3 and R4: two are localised near the wall, and the other two exist at larger $Y$.

Along the convex portion of the bump, the TGVs progressively vanish under the action of the pressure strain, which, close to the wall, reorients the spanwise fluctuations into vertical fluctuations. This explains the negative peak of $\Pi_{33}$ at $(X,r_z,\Delta Y)=(16.7,0.37,0.02)$, and the positive peak of $\Pi_{22}$ at the same scales and positions. Moving downstream, this redistribution mechanism is absent already at $X \approx 17$, as the TGVs do not survive for long (see \S\ref{sec:structures-after}).

The second redistribution mechanism occurs at larger wall distances ($Y \approx 0.2$) and can only be seen for $X \le 17.2$. Here the energy of the streaks advected from the bump tip is partially redistributed towards the cross-stream components and feeds the QSVs. The local negative minimum of $\Pi_{11}$ at $(r_z, \Delta Y) \approx (0.24, 0.1)$ and the local positive maximum of $\Pi_{22}$ and $\Pi_{33}$ at $(r_z, \Delta Y) \approx (0.62, 0.07)$ and $(r_z,\Delta Y) \approx (0.32,0.13)$ at $X=16.9$ support this scenario. For all the diagonal components, the quantity $\Pi_{ij}+P_{ij}$ is positive at these scales and positions, implying that the interaction of the wall cycle with the mean flow is a net source for the velocity fluctuations in the three directions, with the streamwise ones being sustained by production, and the cross-stream ones by redistribution.

In R3 a third redistribution mechanism originates near the wall at $X \approx 16.9$, and acts again to partially reorient streamwise fluctuations into cross-stream ones; see the negative and positive peaks of $\Pi_{11}$ and $\Pi_{33}$ at $X \approx 16.9$, $\Delta Y \approx 0$ and $r_z \approx 0.2$ in the top and bottom panels of figure \ref{fig:pstr_after}. It is associated with the inner layer generated after the convex-concave change of curvature; its intensity and vertical extent grow with the streamwise coordinate (note the streamwise evolution of the corresponding peaks of $\Pi_{ij}$), and at $X \gtrapprox 17.4$ becomes the main redistribution process active in R4.

\begin{figure}
\centering
\includegraphics[trim={0 330 0 100},clip,width=0.9\textwidth]{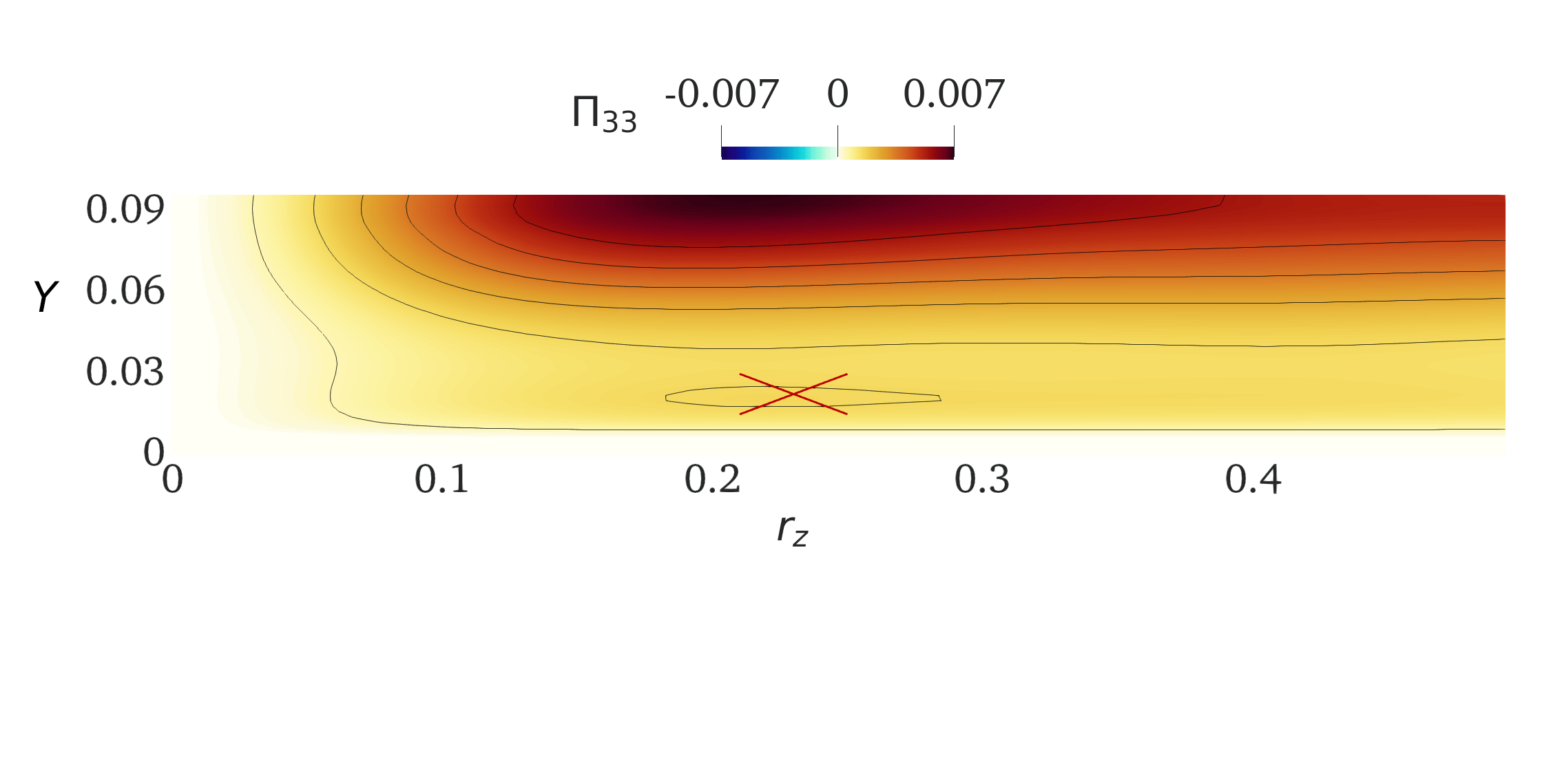}
\includegraphics[trim={0 330 0 100},clip,width=0.9\textwidth]{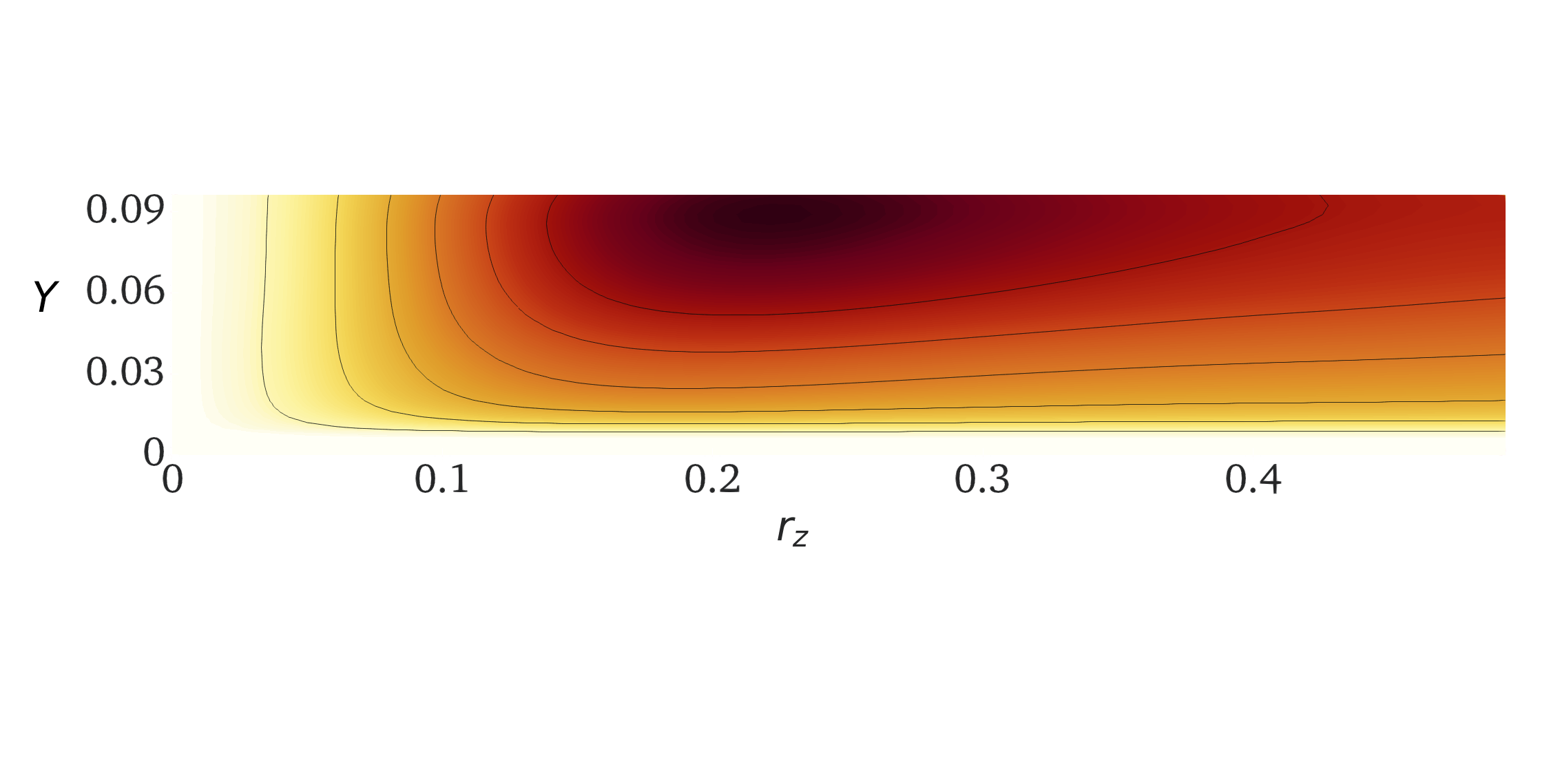}
\caption{Pressure strain term $\Pi_{33}$ is the $Y-r_z$ plane with minimum $C_f$. The top panel is for the bump that induces separation after the tip (considered in this work). The bottom panel is for the bump $G_2$ of \cite{banchetti-luchini-quadrio-2020} that does not induce flow separation. The cross in the top panel highlights the near-wall local peak of $\Pi_{33}$. See caption of figure \ref{fig:ener_after} for contour lines.}
\label{fig:Pi33_bubble}
\end{figure}
Lastly, the fourth redistribution mechanism is at work close to the mean-flow reattachment point and inside the recirculating region. Here $\Pi_{11}$ and $\Pi_{33}$ are positive, with $\Pi_{22}<0$: this is the statistical signature of the flow impinging 
 (see the top panel of figure \ref{fig:Pi33_bubble} for a zoom of $\Pi_{33}$ at $X=17.42$). 
In this region, $\Pi_{33}>\Pi_{11}$, and the local peaks of $\Pi_{22}$ and $\Pi_{33}$ identify a small spanwise scale of $r_z \approx 0.22$. Hence, the pressure--strain preferentially organises the velocity fluctuations in small-scale $w-$structures, similarly to the findings of \cite{chiarini-etal-2022}. 
This mechanism is expected to become more significant when the recirculation is larger and more intense. 
To prove that these $w-$structures are associated with the recirculating bubble, we consider an additional bump that does not induce separation after the bump tip. This second bump has the same upstream part, but an horizontal expansion factor of $2.5$ is applied to the rear part; for further details see bump $G_2$ in \cite{banchetti-luchini-quadrio-2020}. The bottom panel of figure \ref{fig:Pi33_bubble}, plots $\Pi_{33}$ for this bump, at the position of minimum $C_f$ (that in this case is positive as there is no separation). 
The absence of a local peak of $\Pi_{33}$ for $ \Delta Y \rightarrow 0$, confirms that when the flow remains fully attached and does not impinge on the wall, this redistribution mechanism does not occur, and the fluctuations do not organise in $w-$structures.

\section{Scale-space energy transport} 
\label{sec:fluxes}

In this section, the effects of wall curvature on the energy transfers are studied by means of the fluxes of the scale energy $\aver{\delta q^2} = \aver{\delta u_i \delta u_i}$ (repeated indices imply summation), i.e. the sum of the diagonal components of the structure function tensor \citep{marati-casciola-piva-2004, cimarelli-deangelis-casciola-2013, cimarelli-etal-2016}. Differences with the plane channel flow are highlighted. 

The AGKE enable a precise description of the energy transfer through the field lines of the fluxes and their divergence. In fact, the fluxes link scales and positions where production and dissipation are not balanced. Therefore, although energy is not bound to be actually transported along these lines, and there is no causal relation between sources and fluxes, the fluxes explain the different scales and positions where the sources peak, and their field lines help to understand their spatial arrangement. This is not different from other common definitions of fluxes or interscale transfer, which are adopted for example with spectrally-decomposed Reynolds stresses budget \citep[][]{kawata-alfredsson-2018,lee-moser-2019}.

Since the flow is dominated by convection, the mean transport $\psi_{X,ii}^{mean}=\aver{U^* \delta q^2}$ in the $X$ direction overwhelms the other contributions to the vector of fluxes, whose residual scale dependence is minimal. For conciseness, hereafter the repeated indices are omitted and the compact notation $\vect{\Phi} =(\psi_{X},\psi_{Y},\phi_{z})$ is used to signify $\vect{\Phi}_{ii} = (\psi_{1,ii},\psi_{2,ii},\phi_{3,ii})$.
To appreciate the inter-scale transfers, the mean convection can be removed by considering the two-dimensional flux vector $\vect{\Phi}'=(\psi_Y,\phi_z)$ in the $(Y,r_z)$ subspace with $r_x=r_y=0$, evaluated at different $X$ positions. 
In this subspace, the budget equation for $\aver{\delta q^2}$ is written by moving $\partial \psi_X / \partial X$, $\partial \phi_x / \partial r_x$ and $\partial \phi_y / \partial r_y$ to the right-hand side to form an extended source $\xi'$:
\begin{equation}
  \underbrace{ \frac{\partial \psi_Y}{\partial Y} + \frac{\partial \phi_z}{\partial r_z} }_{\vect{\nabla} \cdot \vect{\Phi}'} = 
  \underbrace{ P - D - \frac{\partial \psi_X}{\partial X} - \frac{\partial \phi_x}{\partial r_x} - \frac{\partial \phi_y}{\partial r_y} }_{ \xi'}.
\end{equation}

The flux vector $\vect{\Phi}'$ describes how $\aver{\delta q^2}$ is transferred in the $Y$ direction and across spanwise scales. Its field lines convey directional information, whereas its divergence $\vect{\nabla} \cdot \vect{\Phi}'$ provides quantitative information about the energetic relevance of the fluxes. When $\vect{\nabla} \cdot \vect{\Phi}'>0$, the fluxes are energised by local mechanisms, whereas $\vect{\nabla} \cdot \vect{\Phi}'<0$ indicates fluxes releasing energy to locally sustain $\aver{\delta q^2}$.

\begin{figure}
\centering
\includegraphics[width=1\textwidth]{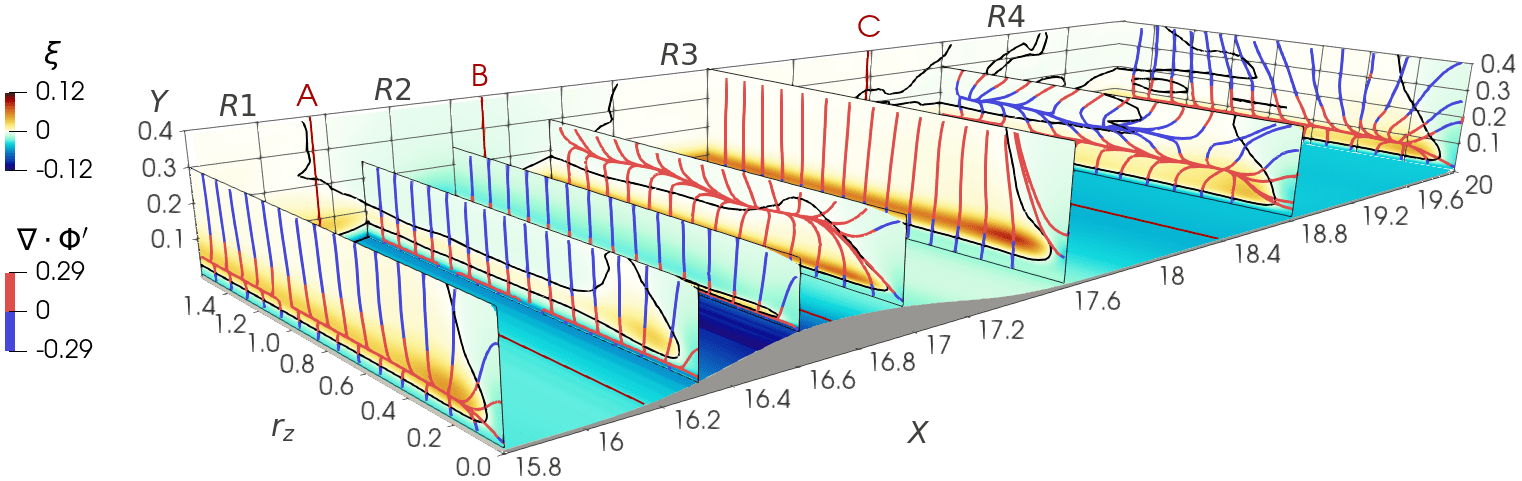}
\caption{Colormap of the complete source term $\xi$ on selected planes (where the black contour line indicates the level $\xi=0$), and field lines of the flux vector $\vect{\Phi}'$, coloured by the flux divergence.}
\label{fig:fluxes_gke}
\end{figure}

Figure \ref{fig:fluxes_gke} shows with colormaps the complete source term $\xi$ defined in Eq.\eqref{eq:source}, together with the field lines of the vector $\vect{\Phi}'$ coloured with its divergence. 
A large source is found at $\Delta Y \approx 0.1$, due to the production mechanisms discussed in \S\ref{sec:prod-before} and \S\ref{sec:prod-after}: it shrinks and weakens before the bump tip at the end of R2, but enlarges and strengthens after the tip in R3 and R4. Sinks, instead, are found in the near-wall region $\Delta Y \rightarrow 0$, at the channel centre $Y \rightarrow h$ and at the smallest scales $r_z \rightarrow 0$, where viscous dissipation dominates and destroys the fluctuating energy. 
The field lines of the flux vector originate from the so-called driving scale range (DSR), i.e. a singularity point of the flux vector generally placed in a source region \citep{cimarelli-deangelis-casciola-2013}. Here fluxes are locally energised, i.e. $\vect{\nabla} \cdot \vect{\Phi}'>0$. The field lines are then attracted to the sinks, where they release energy to locally sustain the fluctuations, and $\vect{\nabla} \cdot \vect{\Phi}'<0$.

In the turbulent plane channel flow, the singularity point is found at ($r_z, \Delta Y) \approx (0.27,0.06)$, or $(r_z^+,\Delta Y^+) \approx (52, 12)$, close to the maximum of $\xi$ (see the plane at $X=15.8$). 
Some field lines vanish at $r_z \rightarrow 0$, with a classic direct energy transfer from larger to smaller scales. Others, instead, are attracted by the wall and by the channel centre, and release energy at both larger and smaller scales, with the coexistence of ascending/descending and direct/inverse energy transfers \citep{cimarelli-deangelis-casciola-2013, chiarini-etal-2021}. 

In R1 the fluxes resemble those of the plane channel. From a quantitative viewpoint, though, the mild adverse pressure gradient in R1 moves the DSR towards the wall and towards smaller scales: for $X=15.8$ the DSR is at $(r_z,\Delta Y)=(0.23,0.05)$, while for $X=16.3$ it moves to $(r_z,\Delta Y) = (0.15,0.01)$.
In R2, instead, the DSR moves towards larger scales, in agreement with the downstream evolution of the near-wall structures discussed in \S\ref{sec:before}; for example the singularity point moves from $r_z=0.15$ at $X=16.3$ to $r_z =0.26$ at $X=16.6$.

Immediately after the bump tip, the adverse pressure gradient in R3 moves the DSR further from the wall and towards larger $r_z$; at $X=17$, the singularity point is at $(r_z,\Delta Y) \approx (0.6, 0.13)$ or $(r_z^+,\Delta Y^+) \approx (76,18)$. This is, once again, consistent with the streamwise evolution of the structures advected from the bump tip. 
Note that, at these $X$, before releasing energy at $\Delta Y \rightarrow 0$, the lines directed towards the wall are further energised by the positive production of the internal layer (see the map of $P_{11,b}$ in \S\ref{sec:prod-after}) at $\Delta Y \approx 0.1$.

Further downstream in R4, the source related to the internal layer strengthens, and a new DSR appears close to the wall. At $X=18.8$ the previous DSR is at $(r_z,Y) \approx (1.5,0.2)$, while the new DSR is at $(r_z,Y) \approx (0.28,0.1)$. 
Here the field lines indicate for $Y \ge 0.2$ an ascending transfer of energy towards the channel centre, and a descending transfer for $ Y \le 0.1$. For intermediate positions, i.e. $0.1 \le Y \le 0.2$, the field lines connect the top and bottom singularity points. The colour of the field lines confirms that the fluxes are mainly energised by the new DSR, in agreement with the larger intensity of the associated production mechanism (see \S\ref{sec:prod-after}).

Later downstream, the intensity of the new DSR increases, while the previous DSR at larger $Y$ progressively weakens and eventually vanishes. At $X = 22$, the effect of the bump disappears: the fluxes recover those on the plane wall, and are fully sustained by the new DSR.

The above description may be useful for its modelling implications in Large Eddy Simulations (LES). In LES, scales larger than the filter length scale (or local grid size) $\Delta$ are resolved, while those smaller than $\Delta$ (subgrid motions) are modelled. Before selecting the model and $\Delta$, one may want to identify the cross-over scale $\ell_{cross}$ that separates the smaller scales dominated by forward energy transfer from the larger scales dominated by reverse energy transfer. When $\Delta < \ell_{cross}$, the subgrid motions are dissipative and models based on the eddy-viscosity assumption can be employed. When $\Delta > \ell_{cross}$, instead, more sophisticated modelling approaches are needed, as the energy of the subgrid motions is fed into larger scales.
A good estimate of the cross-over scale in the spanwise direction, $\ell_{cross,z}$, is the smallest scale $r_z$ at which $\xi=0$. For smaller $r_z$, indeed, $\xi<0$ for all $Y$, which implies a net energy sink and therefore dissipative motions. As the near-wall turbulence evolves along the bump, $\ell_{cross,z}$ is affected by the wall shape and leads to a more strict requirement for spanwise grid resolution, or modelling approach, in the downstream side of the bump. According to our data, the evolution of $\ell_{cross,z}$ with $X$ resembles that of the characteristic spacing of the streaks: the value of $\ell_{cross,z}$ increases over the upstream side of the bump, being $\ell_{cross,z} = 0.062$ or $\ell_{cross,z}^+ = 11$ at $X=15.8$ (region R1) and $\ell_{cross,z} = 0.073$ or $\ell_{cross,z}^+ = 26$ at $X=16.6$ (region R2), and decreases over the downstream side of the bump to $\ell_{cross,z} = 0.053$ or $\ell_{cross,z}^+ = 7$ at $X=17$ (region R3), progressively recovering the channel flow value in R4.

Phenomena such as flow separation, flow recirculation and flow impinging also set strong grid resolution requirements, particularly in the streamwise direction \citep{kuban-etal-2012, mollicone-etal-2018}.

\section{Concluding discussion} 
\label{sec:conclusions}

\begin{figure}
\centering
\includegraphics[width=1\textwidth]{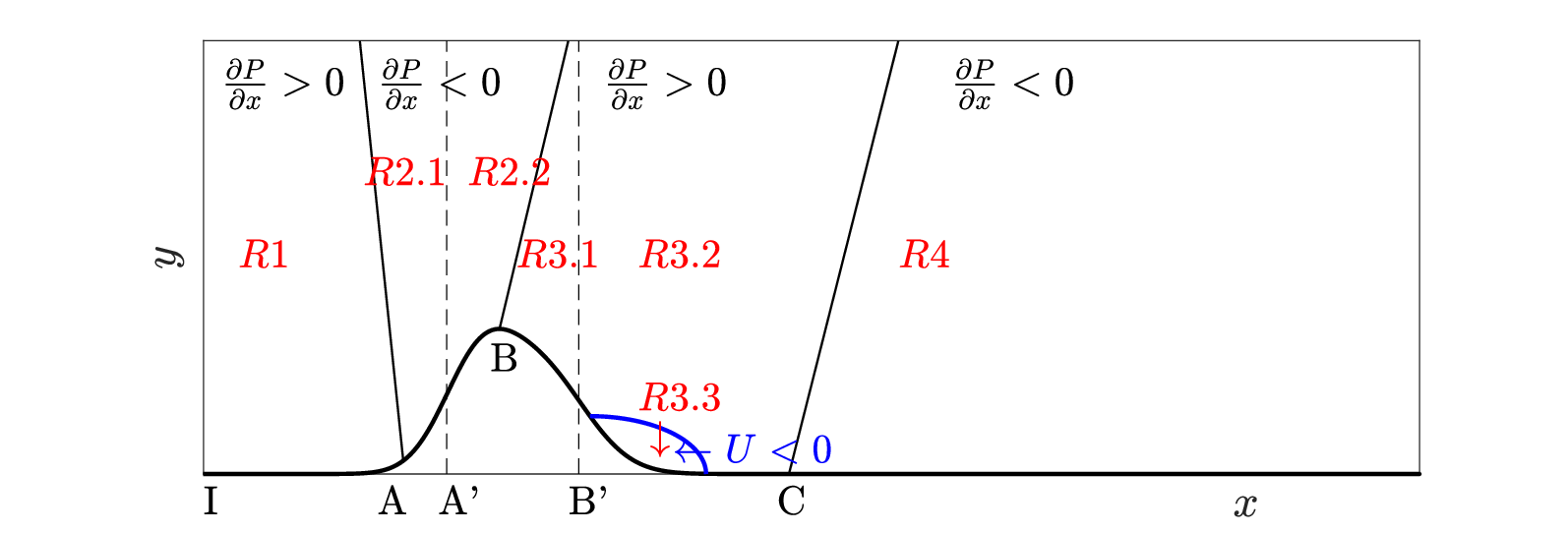}
\caption{Sketch of the different regions in the flow over a bump. The oblique lines identify regions with mean pressure gradient of alternate sign. The vertical dashed lines locate the streamwise position where curvature changes sign. The thick blue line delimits the possible recirculating region.}
\label{fig:sketch}
\end{figure}

In this work the effect of a localised mild curvature of the wall on the structure of near-wall turbulence and its sustaining mechanism is described. 
The study builds upon a DNS database produced by \cite{banchetti-luchini-quadrio-2020} for a turbulent channel flow where a small-height bump on one wall introduces a sequence of concave-convex-concave changes of curvature. 
The two-points statistical analysis employs the anisotropic generalised Kolmogorov equations, or AGKE \citep{gatti-etal-2020}, i.e. budget equations for each component of the structure function tensor, considering simultaneously the space of scales and the physical space. 
Besides a detailed scale-space characterisation of near-wall turbulence in the different regions of the flow, in this work we provide, for the first time in such configuration, statistical evidence of Taylor--G\"ortler vortices over the upstream concave portion of the bump. 

The Reynolds number considered in this work, albeit rather low, is representative of interesting applications. For example, the flow in human arteries can reach similar values of $Re$, especially in presence of severe stenoses in the internal carotid artery \citep{berger-jou-2000}, and poses interesting questions regarding turbulence modelling \citep{lui-etal-2020}. 
The global scenario is expected to be qualitatively unchanged at higher $Re$. 
Although the size of the recirculation region, the intensity of the production of turbulent fluctuations over the bump tip, the size of the flow structures, and the reattachment point of the separating shear layer depend on the Reynolds number based on the bump height \citep{kahler-etal-2016}, the single- and two-points statistical features of the flow remain qualitatively similar. This was recently confirmed by \cite{mollicone-etal-2017, mollicone-etal-2018}; they studied by DNS the turbulent flow past a (larger) bump and found that the main flow features are indeed substantially unaltered for $2500 \le Re_b \le 10000$.
 
Changes of curvature and the ensuing local modifications of the streamwise pressure gradients have been linked to changes of the turbulent near-wall cycle, its typical structures (namely the low-speed streaks and the quasi-streamwise vortices) and the related mechanisms of energy production, dissipation and redistribution.
It should be noted that the bump considered here protrudes from the wall only slightly: its height is comparable to the average wall-normal position of the streaks, and lower than that of the quasi-streamwise vortices. Hence, the bump directly interacts with the streaks, but its effects on the QSVs are indirect. The whole near-wall turbulence cycle is affected, together with its statistical description, which includes energy production and redistribution.

A primary effect of the small bump is to create a streamwise modulation of the pressure gradient, which alternates between favourable and adverse.
Based on its sign, and in combination with the wall curvature, four regions can be defined in the flow; they are shown schematically in figure \ref{fig:sketch}. The upstream side of the bump is divided into regions R1 and R2, with adverse and favourable pressure gradient respectively. Depending on the wall curvature, R2 is further divided into two subregions: R2.1 with concave curvature and R2.2 with convex curvature. Regions R3 and R4 include the downstream side of the bump, with adverse and favourable pressure gradient. R3 can be split into R3.1 and R3.2, with convex and concave curvature. In R3 a recirculation may take place, depending on the size and shape of the bump; the recirculation area is referred to as R3.3.

In R1 the adverse pressure gradient enhances the streamwise and vertical fluctuations, yielding a local peak of the turbulent kinetic energy. 
In R2, instead, the mean flow accelerates, and the turbulent activity decreases, until, close to the bump tip, the energy flow reverses, i.e. from the fluctuating to the mean field. 
As a result, the streaks of streamwise velocity weaken, their characteristic spanwise spacing increases and their distance from the wall decreases. 
Moreover, the non-uniform $V$ component enables the mean field to sustain vertical fluctuations directly, feeding the quasi-streamwise vortices. However, since QSVs are typically higher than the bump, the production of vertical fluctuations is weak. 
In R2.1, where the curvature is concave, the near-wall velocity fluctuations show a tendency to produce spanwise structures with a characteristic spanwise scale of $r_z \approx 0.3 $, which are interpreted as the initial stage of developing Taylor--G\"ortler vortices.
As far as we are aware, this is the first time that such structures are detected in the turbulent flow over a bump. The limited streamwise extent of the upstream concave portion of a bump prevents their complete development, and the noisy turbulent background makes them hardly visible in an instantaneous snapshot.
The dynamics of these structures is dominated by pressure--strain, and their generation is accompanied by an intense redistribution of energy: close to the wall streamwise and (to a lesser extent) vertical fluctuations are reoriented into spanwise ones. As expected by stability arguments, the Taylor--G\"ortler vortices grow under the influence of the concave wall, but are annihilated as soon as the wall turns convex in R2.2. Here the main actor is again the redistribution operated by pressure--strain, which reorients their near-wall spanwise fluctuations into vertical ones.

In R3, after the bump tip, the velocity streaks strengthen again, and progressively recover their unperturbed characteristic spanwise spacing. Here a new production mechanism is at work, draining energy from the mean flow to feed streamwise fluctuations, at all scales and positions, while vertical energy goes from the fluctuating to the mean field. 
Unlike upstream, in R3.2 the concave curvature induces only a small perturbation to the flow, and new Taylor--G\"ortler vortices are not observed. 
The recirculation in R3.3 is rather limited, owing to the small height of the bump. Its dynamics is driven again by pressure--strain, which acts near the impingement area to organise the near-wall fluctuations into small-scale spanwise structures.
Eventually, in R4 the near-wall cycle slowly recovers its canonical state over a plane wall.

The analysis of the energy production and of the scale-space energy transfers may be useful from a modelling perspective. In some regions of the flow, the production terms $P_{11}$ and $P_{22}$ for streamwise and vertical fluctuations are negative. This is a serious challenge for turbulence closures, as, for example, RANS models based on the classic mixing length approach would be unable to represent it \citep{cimarelli-etal-2019-negative}. Moreover, based on the scale-space energy transfers and on the energy source term, we have shown the influence of the localised change of curvature on the spanwise cross-over scale $\ell_{cross,z}$, that is needed for selecting the proper filter scale in LES simulations. A good candidate for $\ell_{cross,z}$ is the smallest spanwise scale for which the source term $\xi$ is zero. By examining the streamwise evolution of $\ell_{cross,z}$, our data show that the requirements for the spanwise grid resolution and/or subgrid modelling approach are more strict over the downstream side of the bump.

Statistical quantities presented in this paper are available online at the following DOI: \url{https://doi.org/10.5281/zenodo.7879911}.

\backsection[Declaration of Interests]{The authors report no conflict of interest.}
\backsection[Author ORCIDs]{

Davide Selvatici, https://orcid.org/0000-0003-3016-1123;

Maurizio Quadrio, https://orcid.org/0000-0002-7662-3576;

Alessandro Chiarini, https://orcid.org/0000-0001-7746-2850.}

\bibliographystyle{jfm}

\end{document}